\documentclass[conference]{IEEEtran}
\IEEEoverridecommandlockouts
\usepackage{cite}
\usepackage{amsmath,amssymb,amsfonts}
\usepackage{algorithmic}
\usepackage{graphicx}
\usepackage{textcomp}
\usepackage{xcolor}

\usepackage{nameref}

\usepackage[utf8]{inputenc}
\usepackage{float}
\usepackage{subfig}

\def\BibTeX{{\rm B\kern-.05em{\sc i\kern-.025em b}\kern-.08em
    T\kern-.1667em\lower.7ex\hbox{E}\kern-.125emX}}

\makeatletter
\def\ps@IEEEtitlepagestyle{%
  \def\@oddfoot{\mycopyrightnotice}%
  \def\@evenfoot{}%
}
\def\mycopyrightnotice{%
  {\footnotesize The copyright belongs to me!\hfill}
  \gdef\mycopyrightnotice{}
}
\def\ps@IEEEtitlepagestyle{%
  \def\@oddfoot{\mycopyrightnotice}%
  \def\@evenfoot{}%
}
\def\mycopyrightnotice{%
  {\footnotesize 978-1-5386-8375-0/18/\$31.00 \copyright 2018 IEEE \hfill}
  \gdef\mycopyrightnotice{}
}
    
\begin{document}

%

\title{Characterizing Twitter Interaction during \\
COVID-19 pandemic using \\
Complex Networks and Text Mining}

\author{
\IEEEauthorblockN{Josimar Edinson Chire Saire}
\IEEEauthorblockA{\textit{Institute of Mathematics and Computer Science (ICMC)} \\
\textit{University of São Paulo (USP)}\\
São Carlos, SP, Brazil \\
jecs89@usp.br}
}

\maketitle

\begin{abstract}

The outbreak of covid-19 started many months ago, the reported origin was in Wuhan Market, China. Fastly, this virus was propagated to other countries because the access to international travels is affordable and many countries have a distance of some flight hours, besides borders were a constant flow of people. By the other hand, Internet users have the habits of sharing content using Social Networks and issues, problems, thoughts about Covdid-19 were not an exception. Therefore, it is possible to analyze Social Network interaction from one city, country to understand the impact generated by this global issue. South America is one region with developing countries with challenges to face related to Politics, Economy, Public Health and other. Therefore, the scope of this paper is to analyze the interaction on Twitter of South American countries and characterize the flow of data through the users using Complex Network representation and Text Mining. The preliminary experiments introduces the idea of existence of patterns, similar to Complex Systems. Besides, the degree distribution confirm the idea of having a System and visualization of Adjacency Matrices show the presence of users' group publishing and interacting together during the time, there is a possibility of identification of robots sending posts constantly.

\end{abstract}


\section{Introduction}

Nowadays, the use of Social Networks to communicate, share information, thoughts, ideas is very common. Usually, people is creating posts, writing during the day and tagging friends, colleagues, etc. Therefore, all this flow of data can represent the actual status of the citizens. Besides considering pandemic covid-19, users can reflect what they are thinking, feeling in front of the global issue related to the pandemic in their cities, countries. In consequence, this behaviour can be analyzed to monitor the situation of the population, health area as Infomediology studies the behaviour through data and Infovelliance is the application using Computational tools and directed/undirected sources of data. 

Actually, there are many studies related to covid-19 to analyze people's behaviour using Social Networks, in particular Twitter because this Social Network presents facilities to access data and the quantity of data can be representative, i.e. top concerns of users \cite{abd2020top} and focus in countries as scope of study: Italy \cite{de2020infoveillance}, Peru \cite{ChireSaire2020.05.24.20112193} , United Kingdom, Unites States \cite{Wang2020.07.12.20151936}, Mexico( cite{ChireSaire2020.05.07.20094466}, Colombia \cite{ChireSaire2020.07.02.20145425}, United States \cite{Samuel2020.06.01.20119362}, Ghana \cite{ChireSaire2020.08.15.20175810}, France \cite{Chire_Saire_2020}. These studies uses a Data Mining approach and Natural Language Processing techniques to describe and understand the phenomenon in many levels: public health, social, mental health and more. 
By contrast, this flow of data can include misinformation produced intentionally through robots. One approach to represent and analyze this networks of users exchanging data is to use Complex Networks. Complex Networks had many applications in different areas: Physics, Biology and Social Sciences. Then, Complex Networks is a capable representation to study the interaction of users, i.e. \cite{pastorescuredo2020characterizing} studies the top users in Spain using Twitter as source data.

The contribution of this paper:
\begin{itemize}
    \item Select South America as scope and study the covid-19 pandemic influence on this region, open the possibilities of studying this phenomenon and provide a proposal for this analysis, section II )
    \item Introduce Complex Networks to study this phenomenon getting data from Social Networks and find pattern related to Complex Systems.
    \item Find a affordable way to identify network of users with constant flowing of data, beside the possibility of finding robots, fake users through this mechanism, section III.
\end{itemize}

\section{Proposal}
\label{section:proposal}

This paper is analyzing the interaction of South American users where Spanish is official language, through Twitter Social Network. Considering Internet Access and density of population, the capital of each country were selected for the analysis.

\subsection{Dataset}

The dataset is a collection of tweets from 08 March to 11 July using Twitter API, the table Tab. 1 describes the dataset.

\begin{table}[hbpt]
\caption{ Description of Dataset }
\centering
\tiny{
\begin{tabular}{|c|c|c|c|c|c|}
\hline
                   & \textbf{Capital} & \textbf{Latitude} & \textbf{Longitude} & \textbf{Tweets} & \textbf{\begin{tabular}[c]{@{}c@{}}Unique\\  Users\end{tabular}} \\ \hline
\textbf{Argentina} & Buenos Aires     & -34.583           & -58.667            & 3872212         & 749842                                                           \\ \hline
\textbf{Bolivia}   & La Paz           & -16.5             & -68.15             & 134605          & 21850                                                            \\ \hline
\textbf{Chile}     & Santiago         & -33.45            & -70.667            & 3146307         & 309351                                                           \\ \hline
\textbf{Colombia}  & Bogota           & 4.6               & -74.083            & 1688045         & 315794                                                           \\ \hline
\textbf{Ecuador}   & Quito            & -0.217            & -78.5              & 836017          & 102979                                                           \\ \hline
\textbf{Paraguay}  & Asuncion         & -25.267           & -57.667            & 1628146         & 156384                                                           \\ \hline
\textbf{Peru}      & Lima             & -12.05            & -77.05             & 2054415         & 287261                                                           \\ \hline
\textbf{Uruguay}   & Montevideo       & -34.85            & -56.167            & 826405          & 192381                                                           \\ \hline
\textbf{Venezuela} & Caracas          & 10.483            & -66.867            & 4102893         & 305684                                                           \\ \hline
\end{tabular}
}
\end{table}

\subsection{Complex Network Construction}

The Complex Network was created considering the next criterion:

\begin{itemize}
    \item Pick the $N_1$ users with more posts during the period of study
    \item Search @tag\_mentions of users inside of the tweets to find users connection
    \item Find a global list of users and create a set from this elements to avoid duplicated users
    \item Create a global text with text for each country to find the users and count them
    \item Create the edges considering the $N_2$ top users and the set of users with the frequency as weight
\end{itemize}

\section{Experiments and Results}
\label{sec:experiments}

This section explains the performed experiments to describe, analyze and understand the interaction of South American Twitter users. 

\subsection{Firs Experiment}

\begin{figure}[hbpt]
\centerline{
\includegraphics [width=0.50\textwidth]{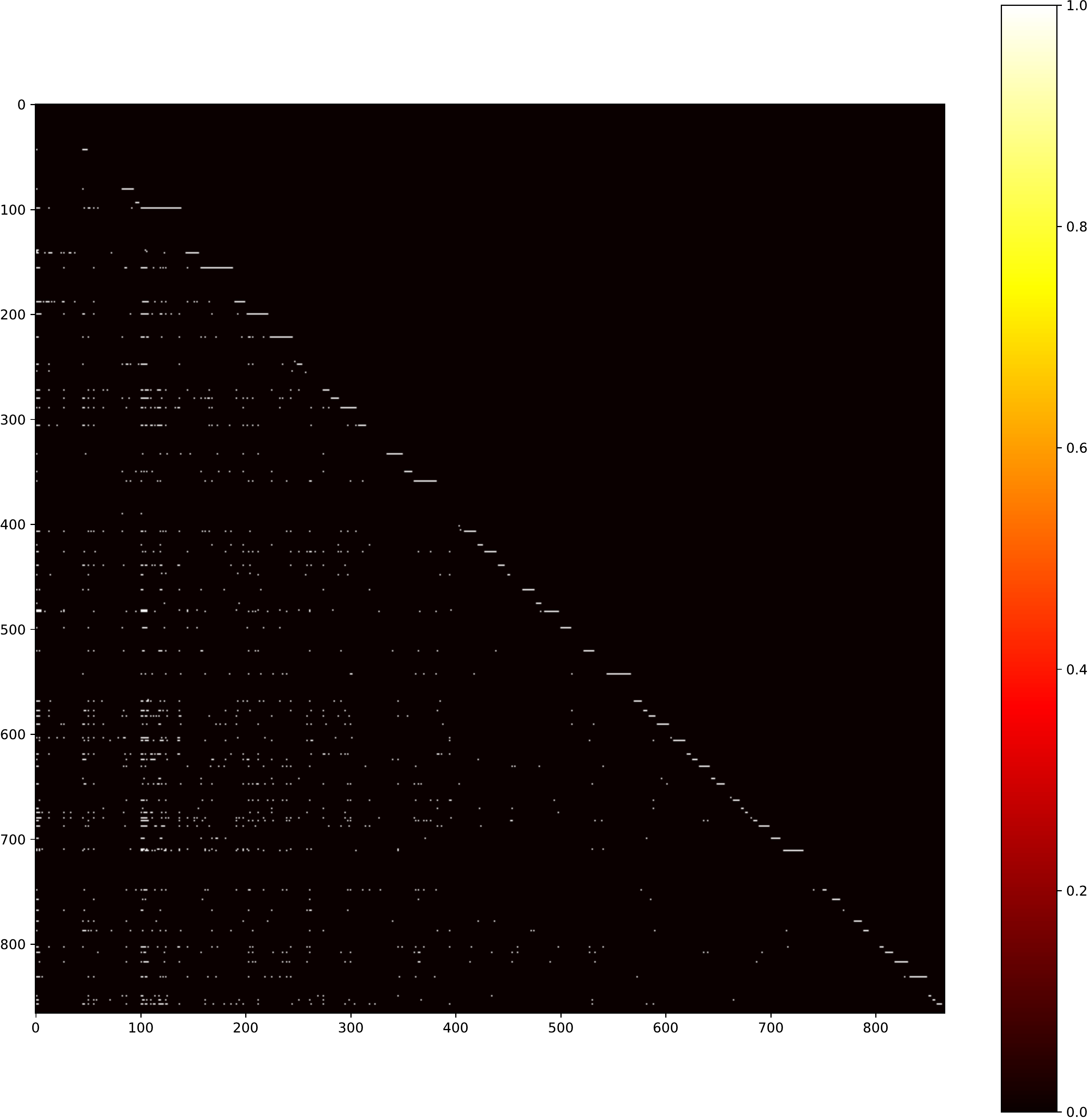}
}
\caption{Peru}
\label{fig:3_1}
\end{figure}

The first experiment used $N_1$ = 500, $N_2$ = 100 and created a directed graph with no associated weights. The graphic \ref{fig:3_1} is presenting the results for Peru country, then it is possible to notice a kind of reticular pattern. Therefore, there is the possibility of existence of users' group.

For the next experiment, the frequency of @tag\_users is considered to set the weight of the edges.

\subsection{Second Experiment}

\begin{figure}[hbpt]
\centerline{
\includegraphics [width=0.50\textwidth]{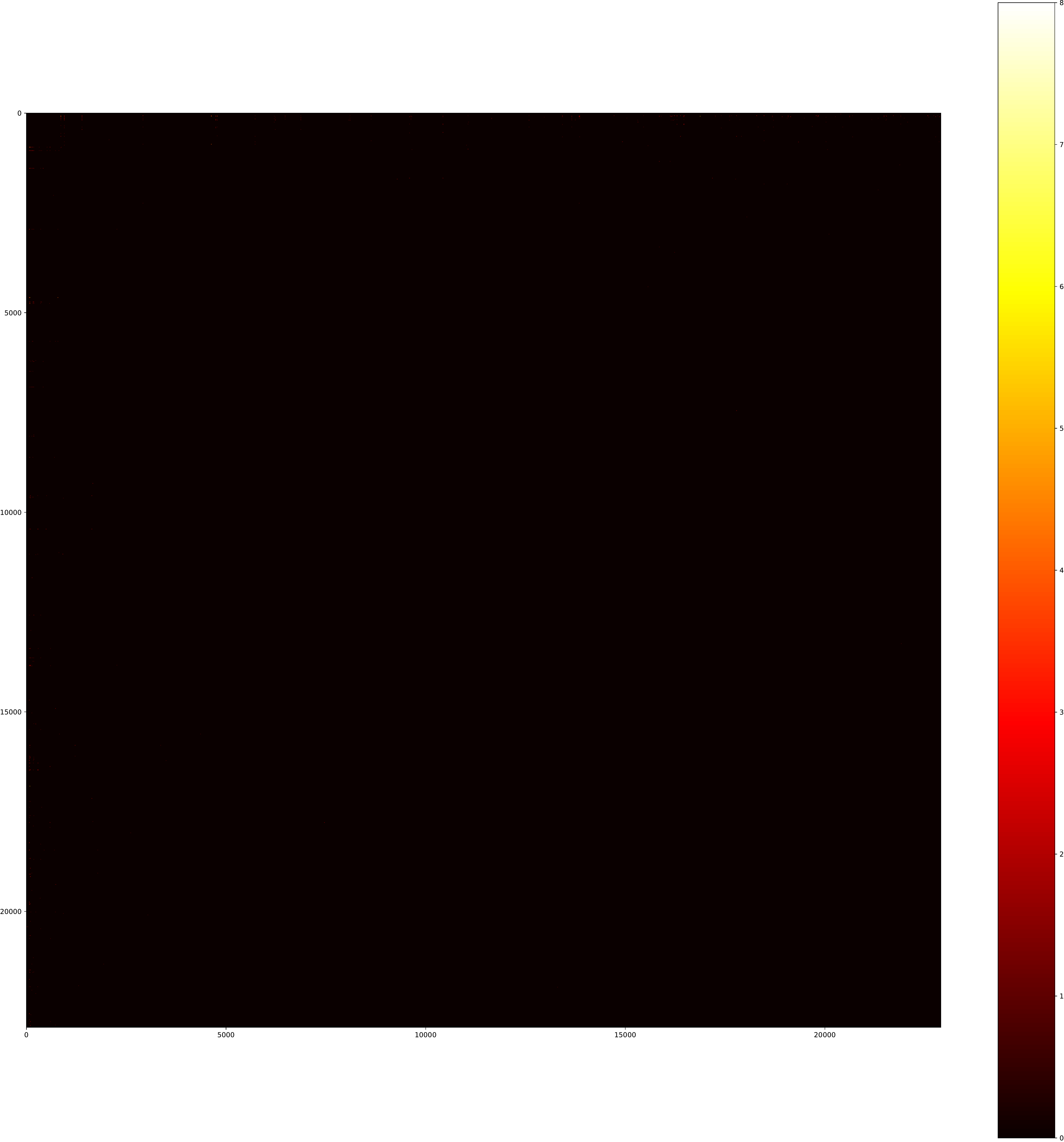}
}
\caption{Peru}
\label{fig:3_11}
\end{figure}

The first experiment used $N_1$ = 2000, $N_2$ = 200 and created a directed graph with no weights related. The graphic \ref{fig:3_11} introduces the result for Argentina, the number of edges for this Complex Network is higher than 20,000 and the weight are very disperse, then a log scale is introduced. In spite of the adaptation it is not possible to perceive the edges or connections between users.

\begin{figure}[hbpt]
\centerline{
\subfloat[Argentina, Bolivia, Chile]{
\includegraphics [width=0.16\textwidth]{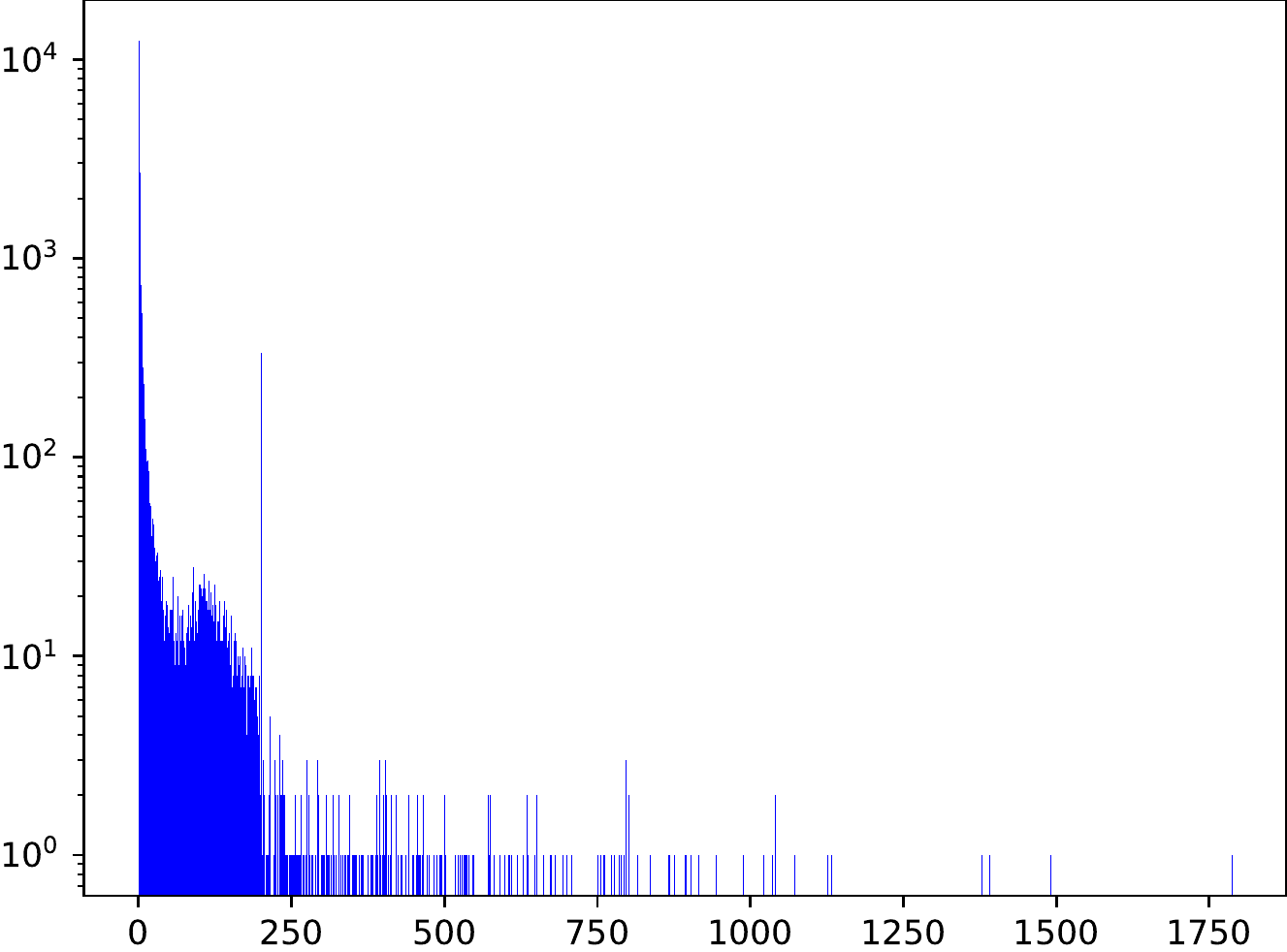}
\includegraphics [width=0.16\textwidth]{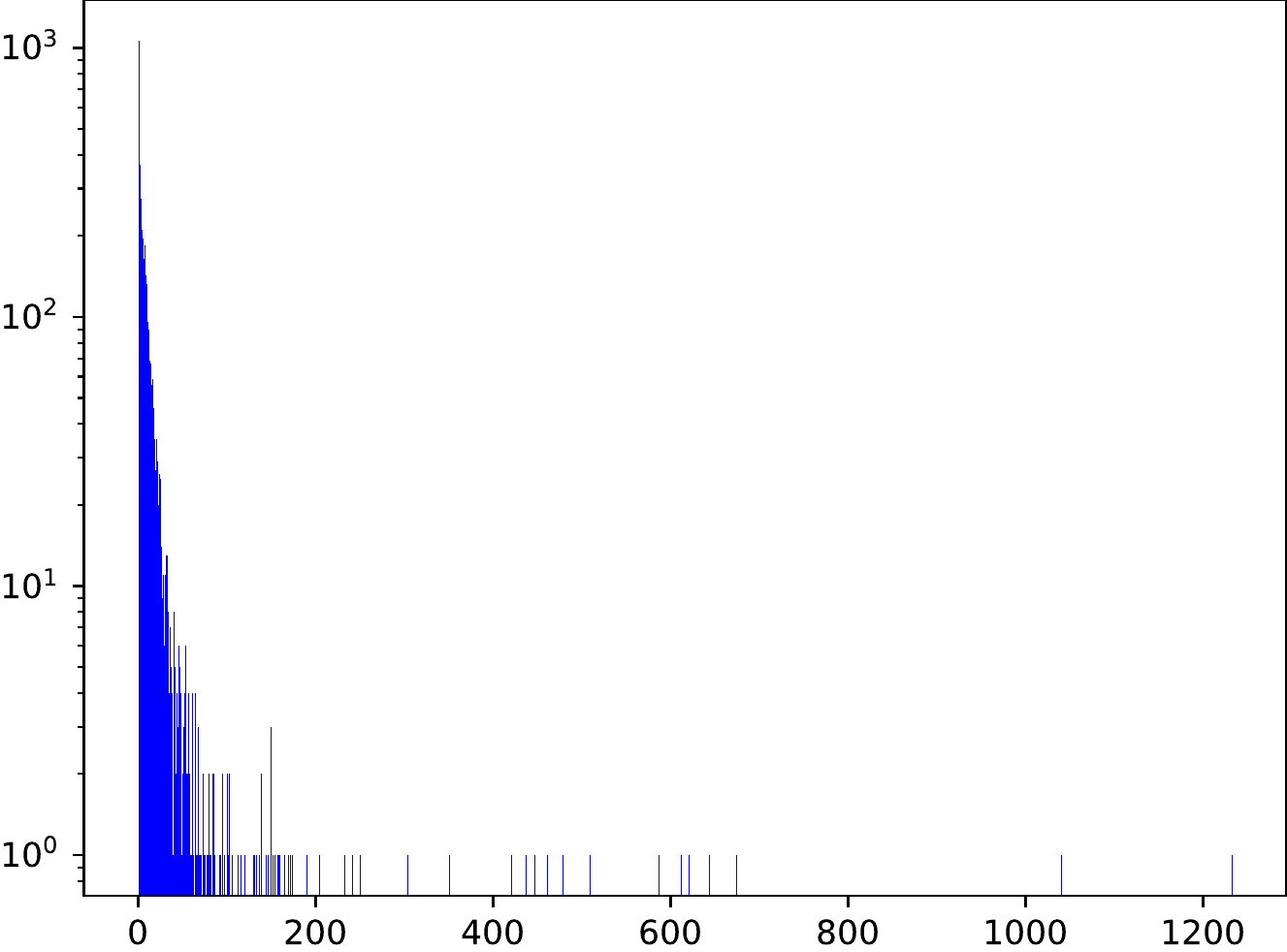}
\includegraphics [width=0.16\textwidth]{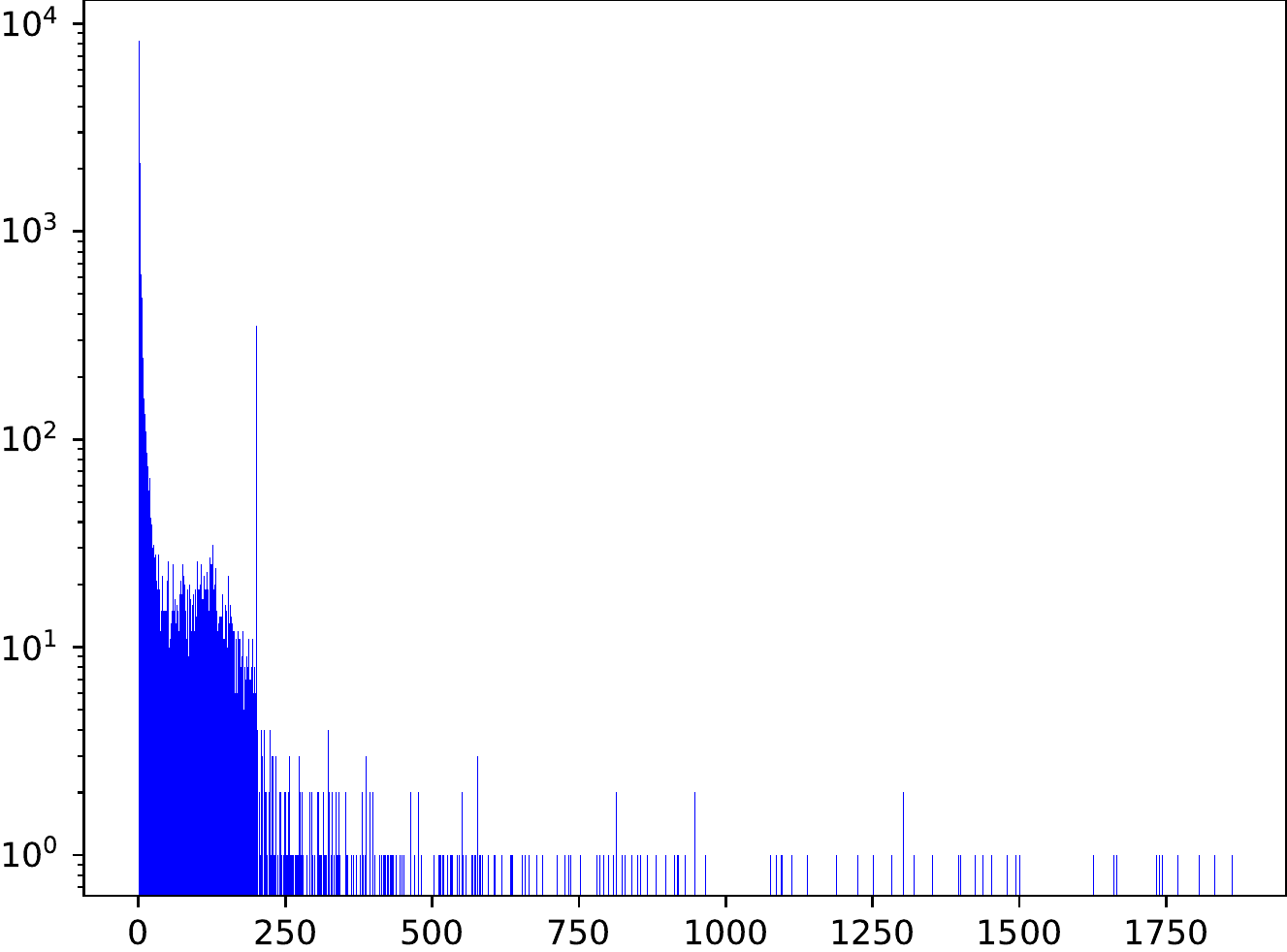}}
}
\centerline{
\subfloat[Colombia, Ecuador, Peru]{
\includegraphics [width=0.16\textwidth]{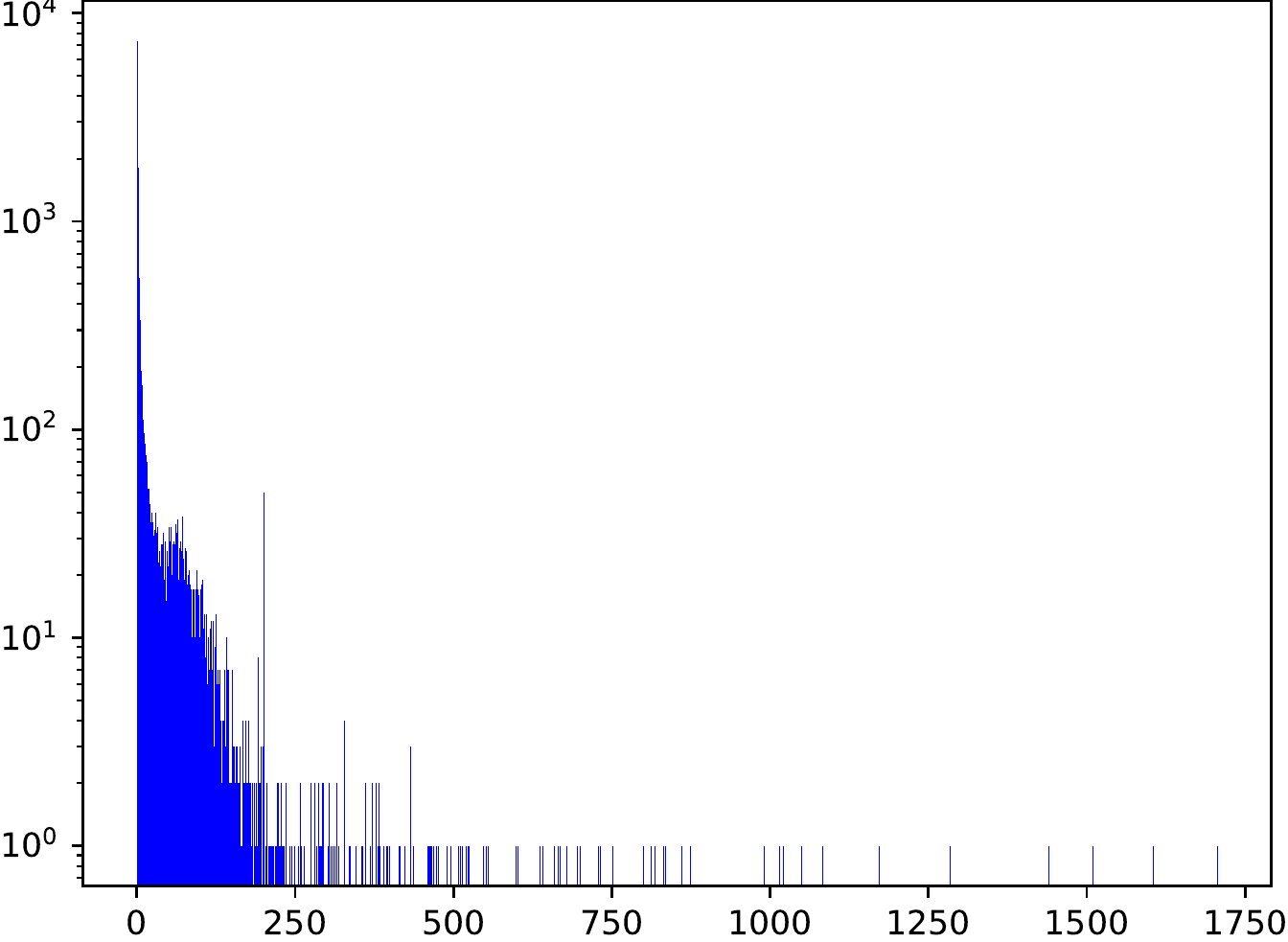}
\includegraphics [width=0.16\textwidth]{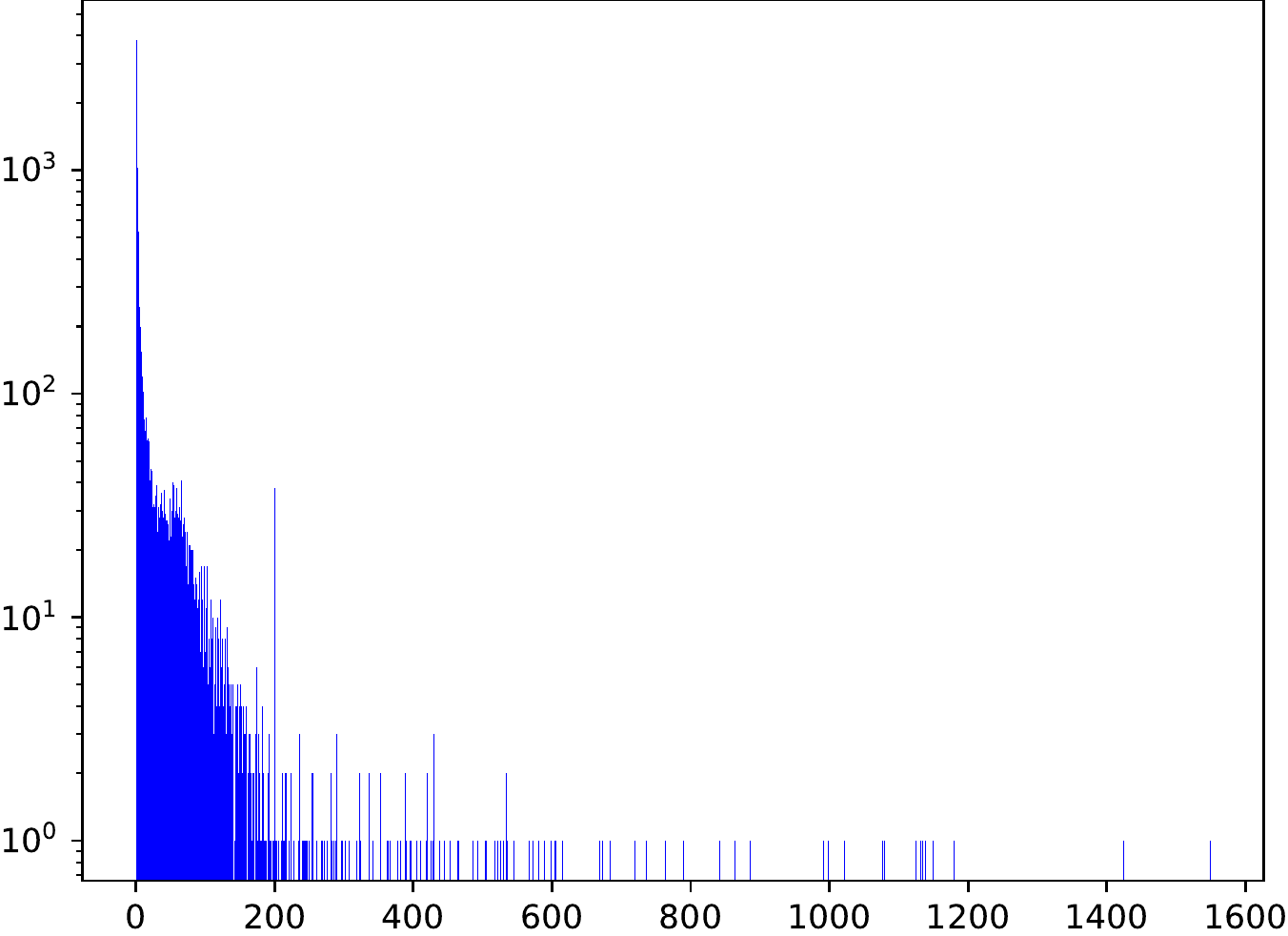}
\includegraphics [width=0.16\textwidth]{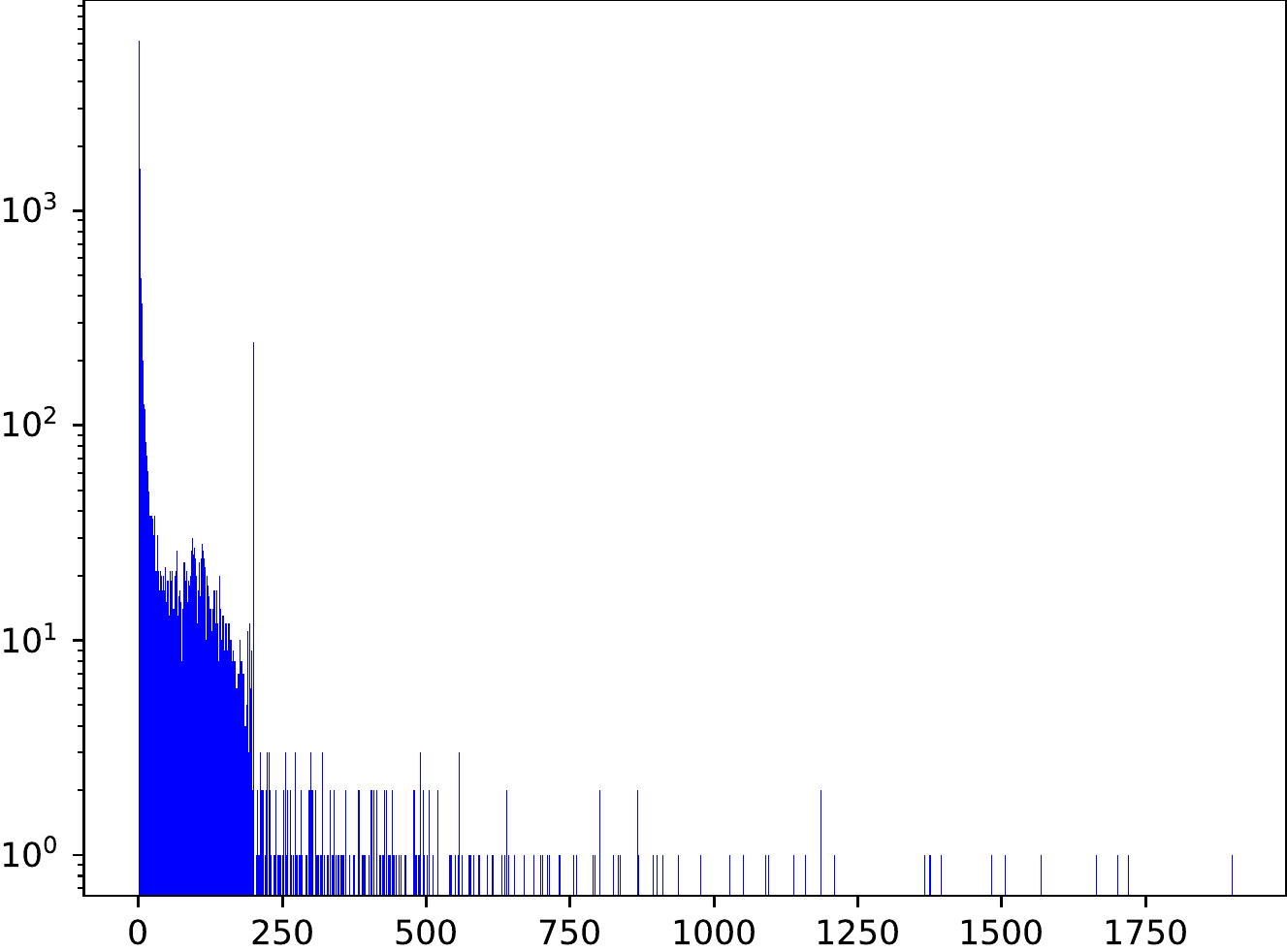}}
}
\centerline{
\subfloat[Paraguay, Uruguay, Venezuela]{
\includegraphics [width=0.16\textwidth]{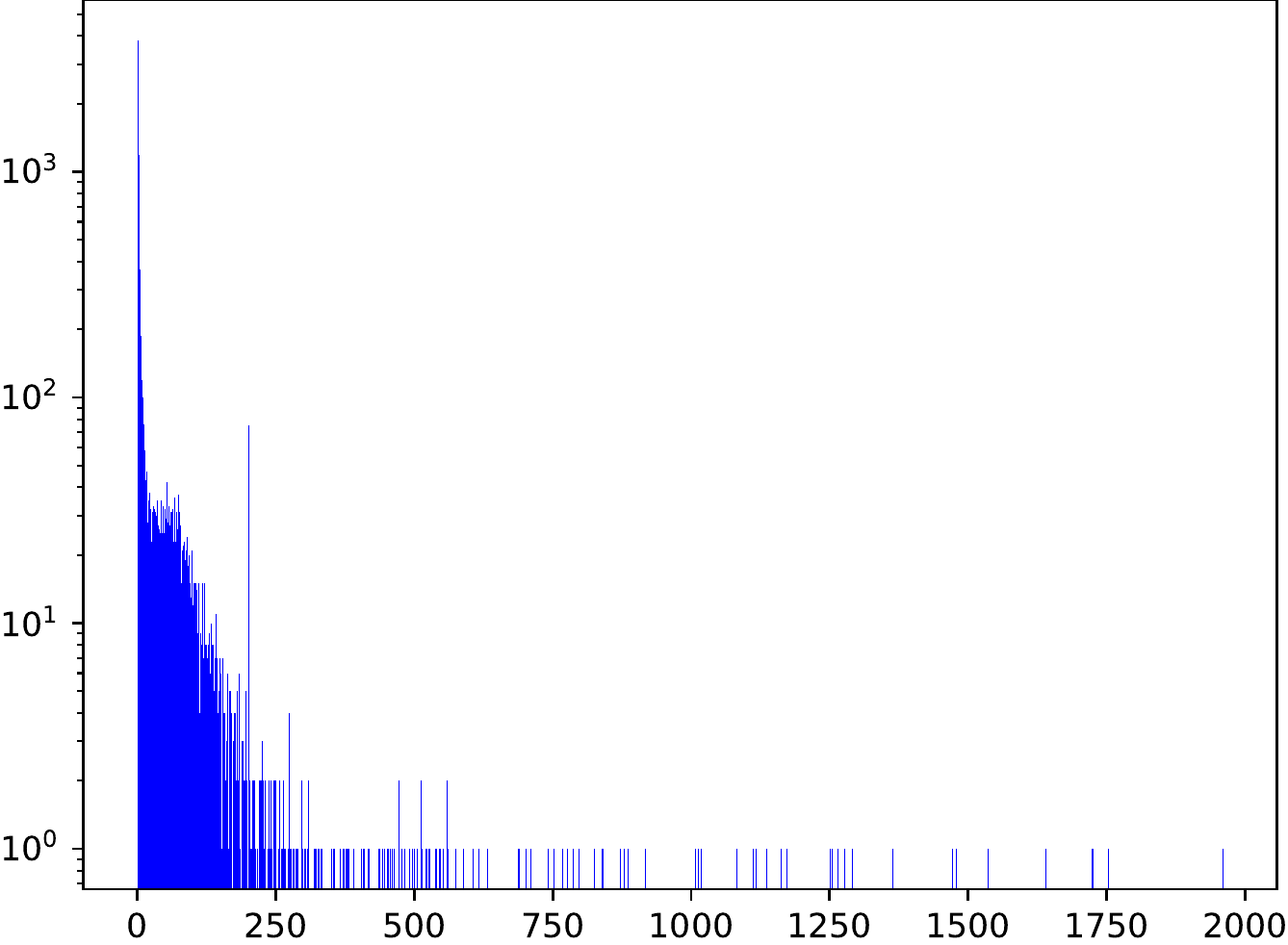}
\includegraphics [width=0.16\textwidth]{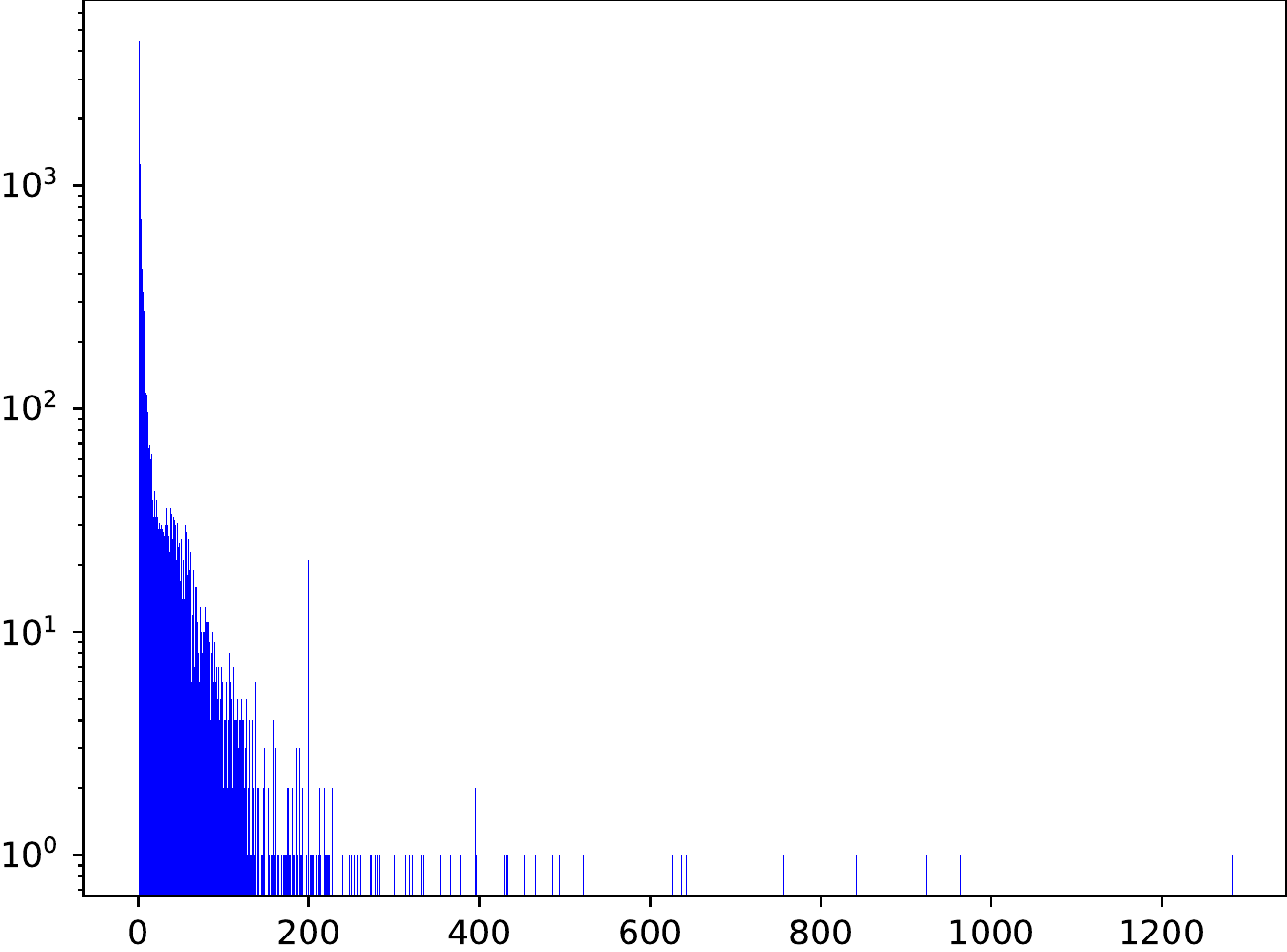}
\includegraphics [width=0.16\textwidth]{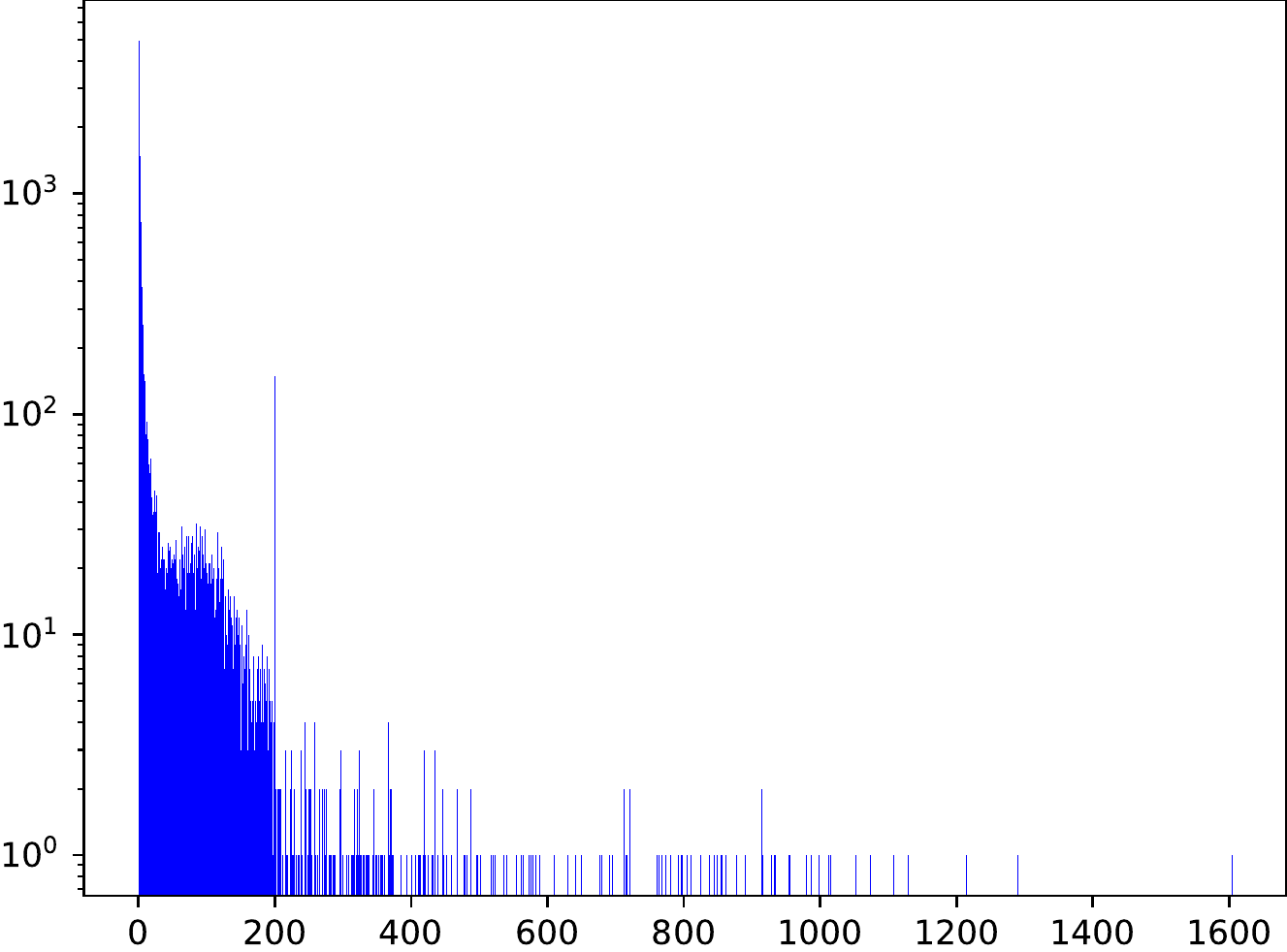}}
}
\caption{South American countries - Degree Distribution}
\label{fig:3_12}
\end{figure}

For the previous reason, a filtering is performed considering the degree distribution(see Figure \ref{fig:3_12}) and only edges with weight higher than 200 are considered. Besides, it is important to notice the presence of a distribution similar to Levy's Distribution.

\begin{figure*}[hbpt]
\centerline{
\subfloat[Argentina, Bolivia, Chile]{
\includegraphics [width=0.28\textwidth]{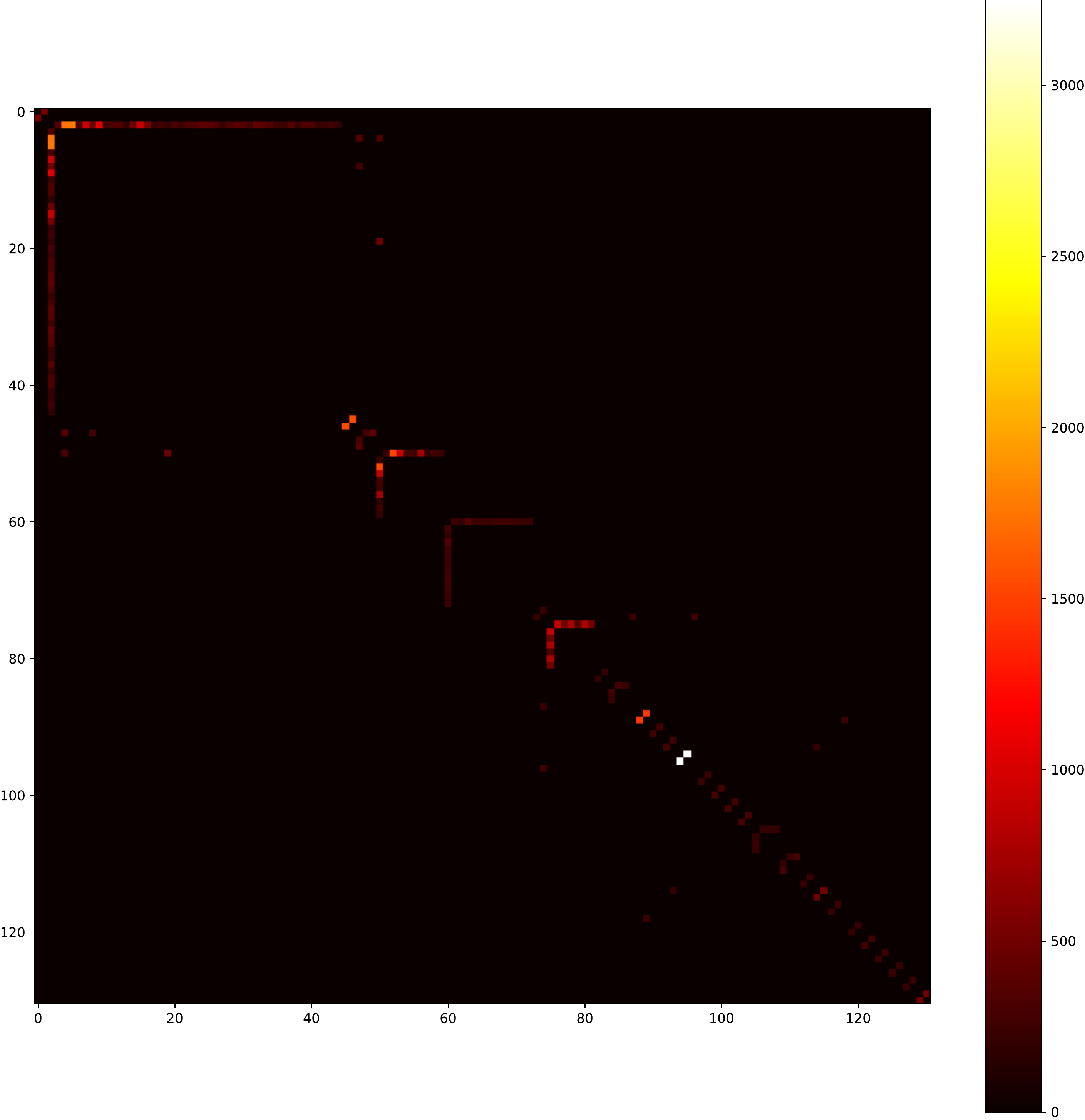}
\includegraphics [width=0.28\textwidth]{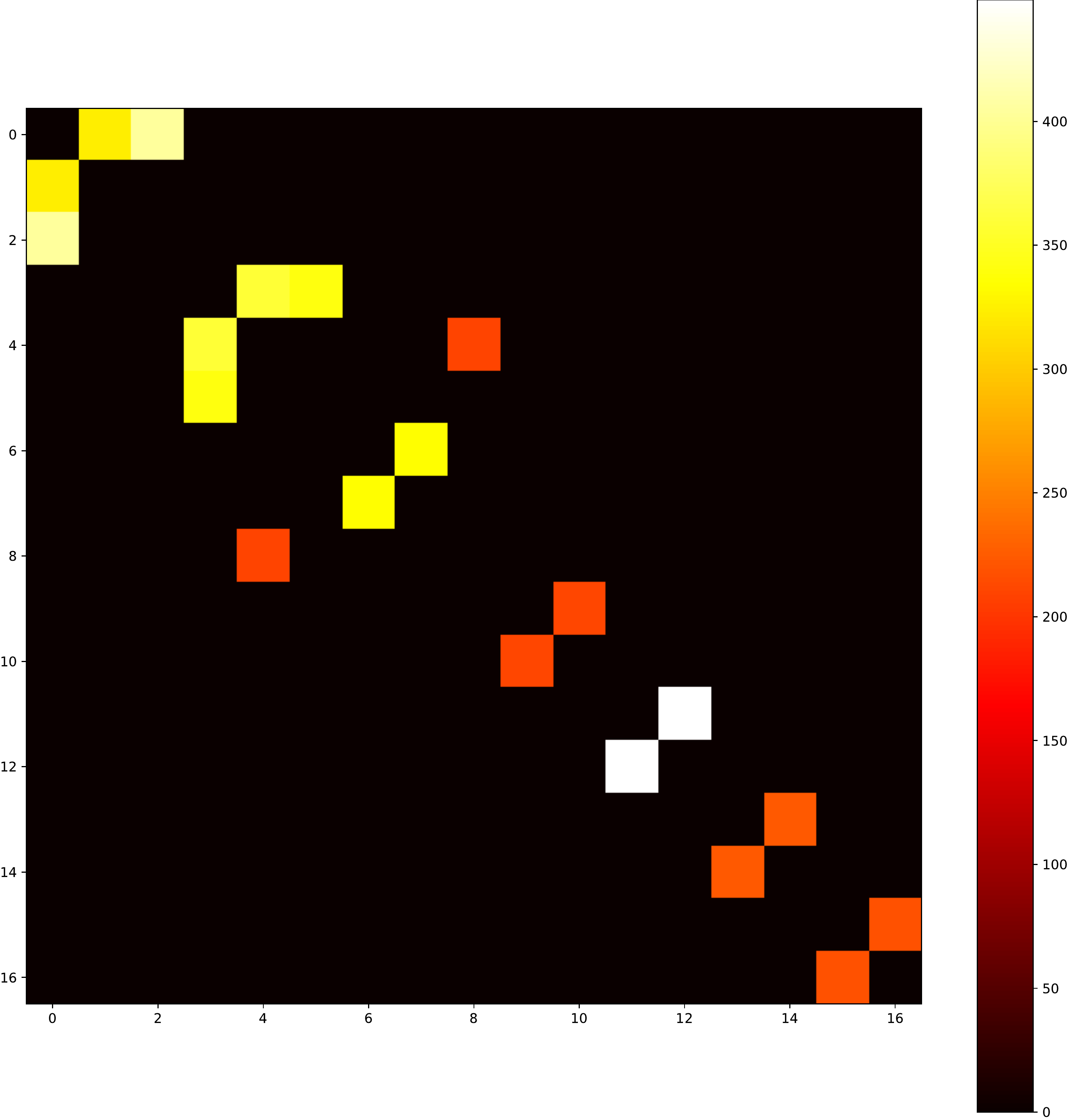}
\includegraphics [width=0.28\textwidth]{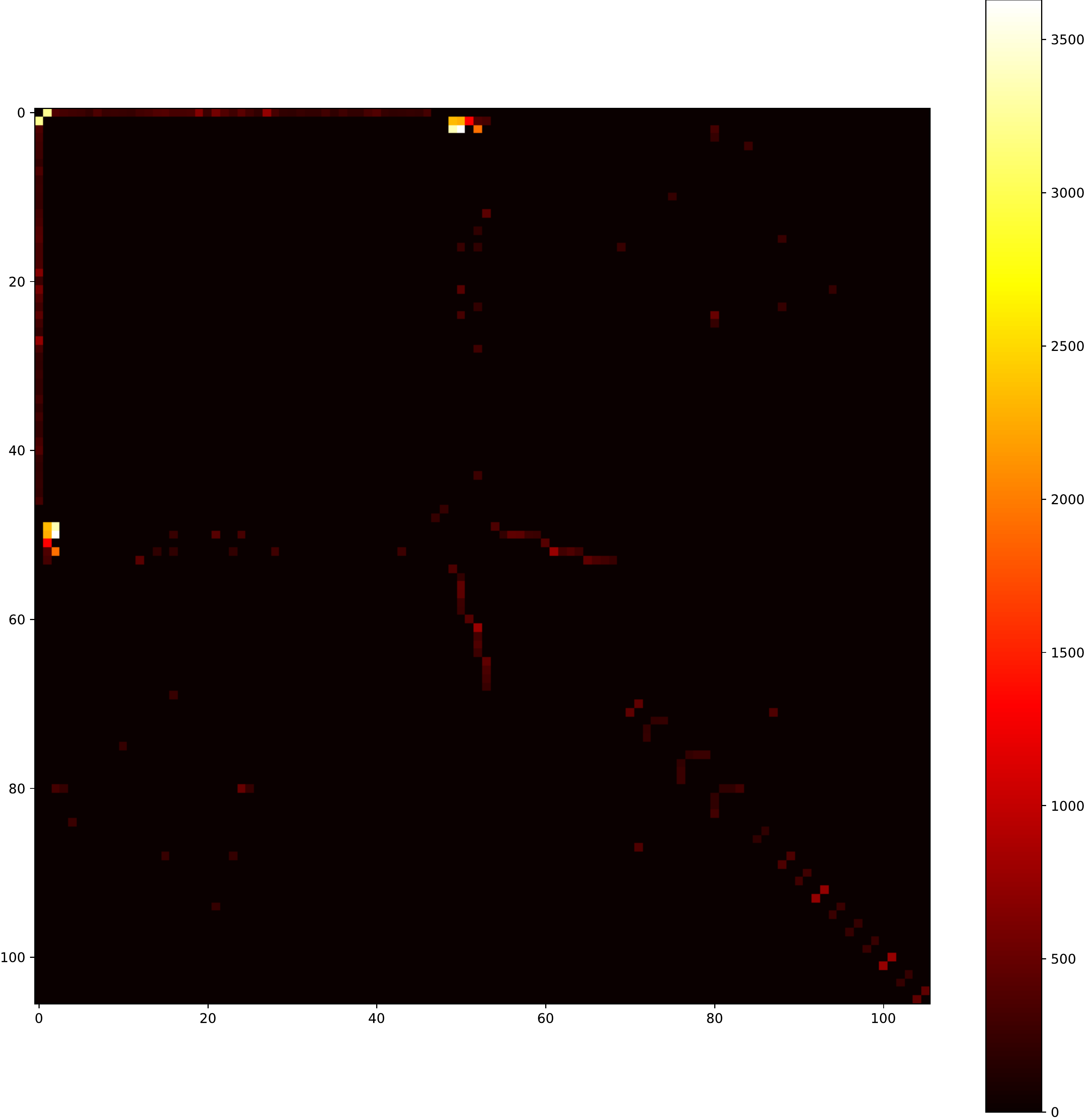}}
}
\centerline{
\subfloat[Colombia, Ecuador, Peru]{
\includegraphics [width=0.28\textwidth]{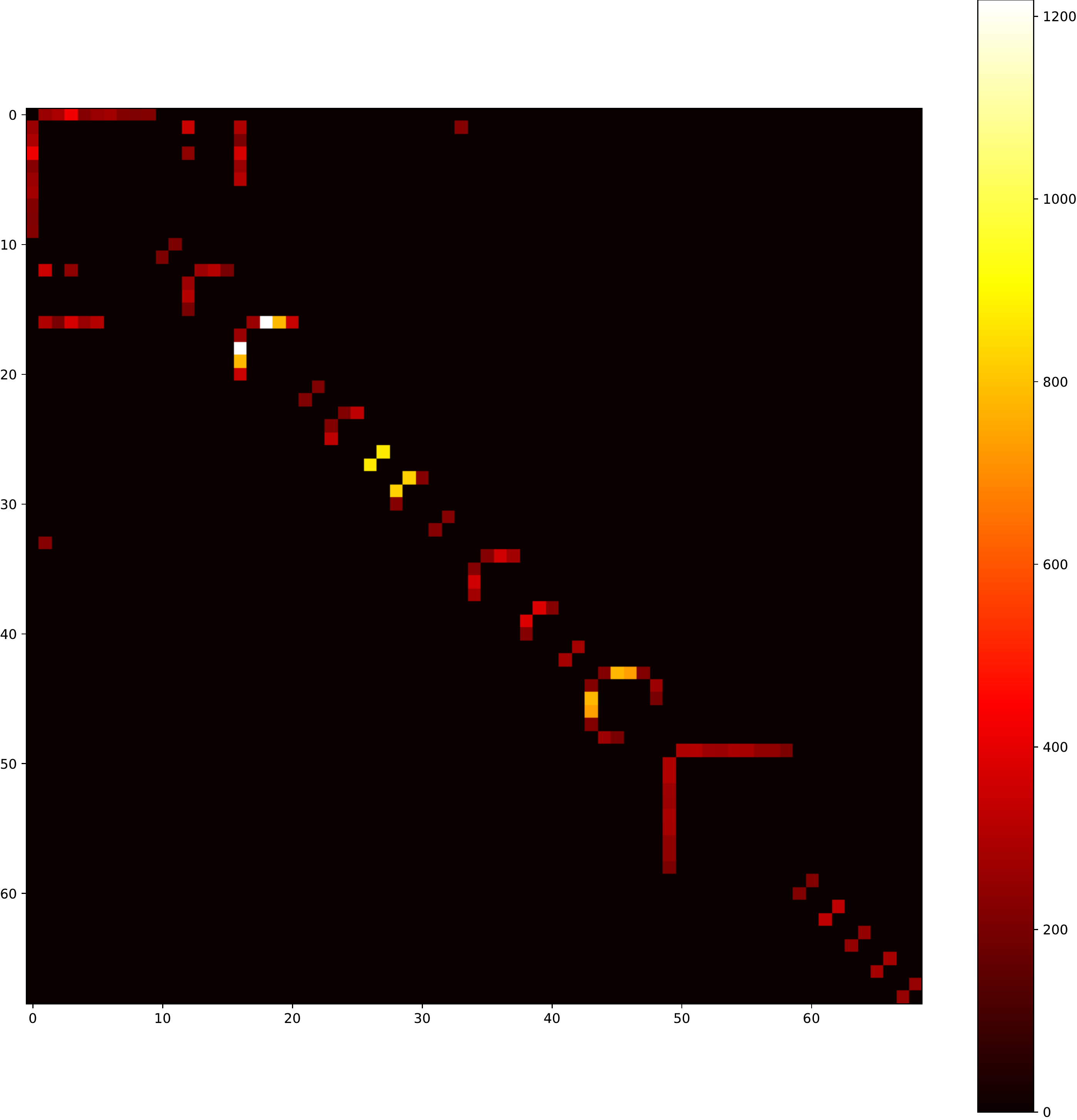}
\includegraphics [width=0.28\textwidth]{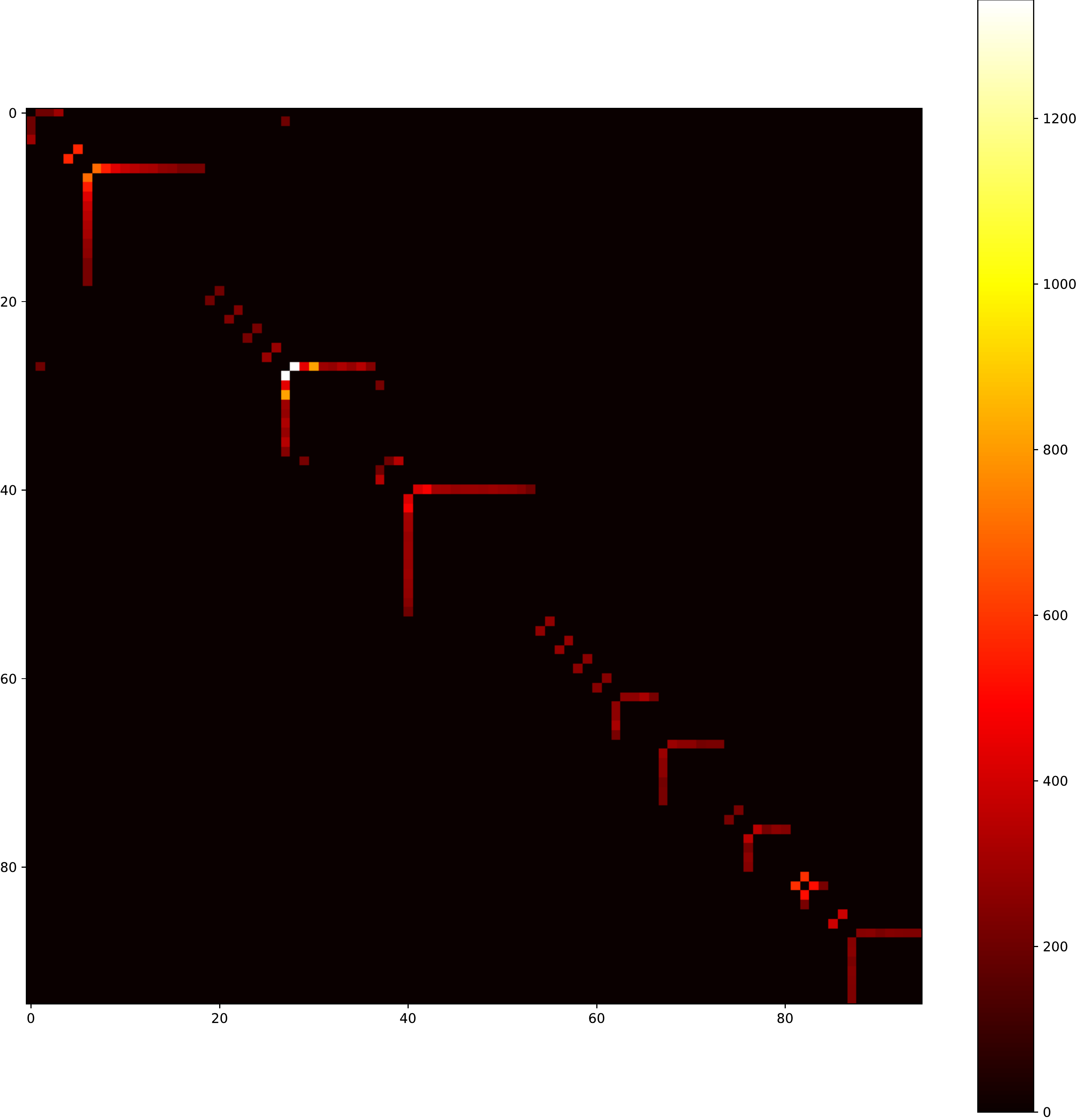}
\includegraphics [width=0.28\textwidth]{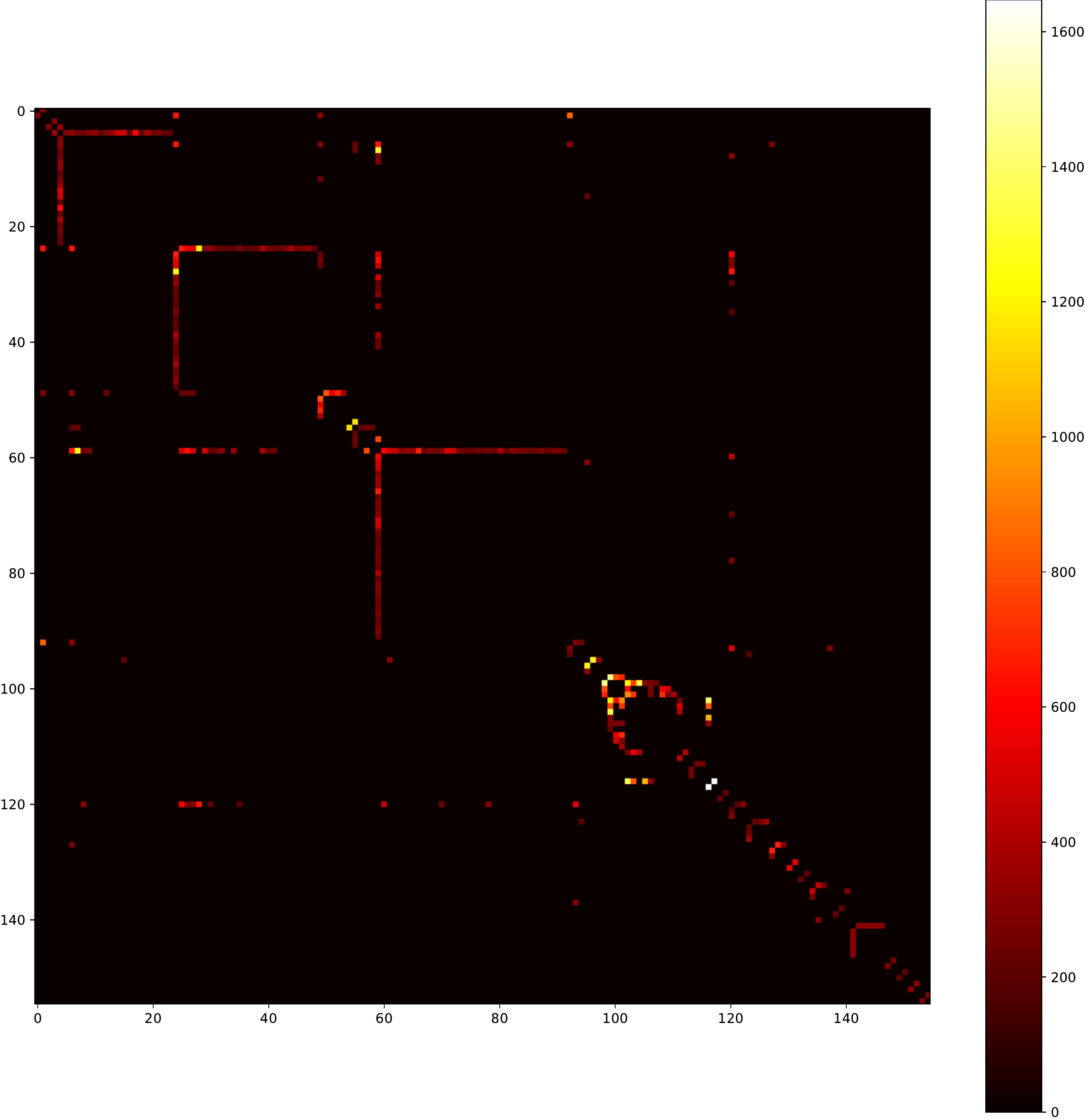}}
}
\centerline{
\subfloat[Paraguay, Uruguay, Venezuela]{
\includegraphics [width=0.28\textwidth]{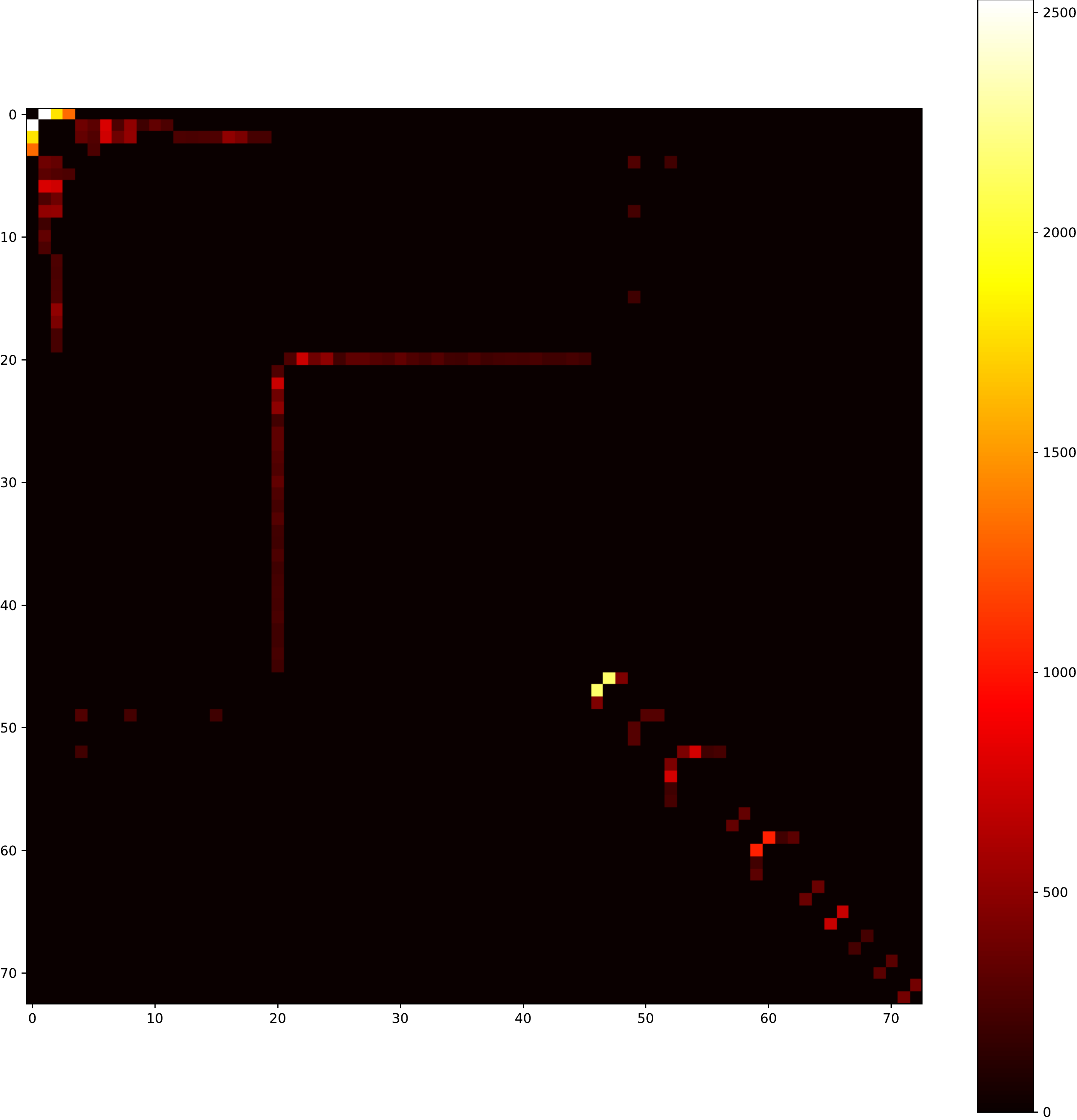}
\includegraphics [width=0.28\textwidth]{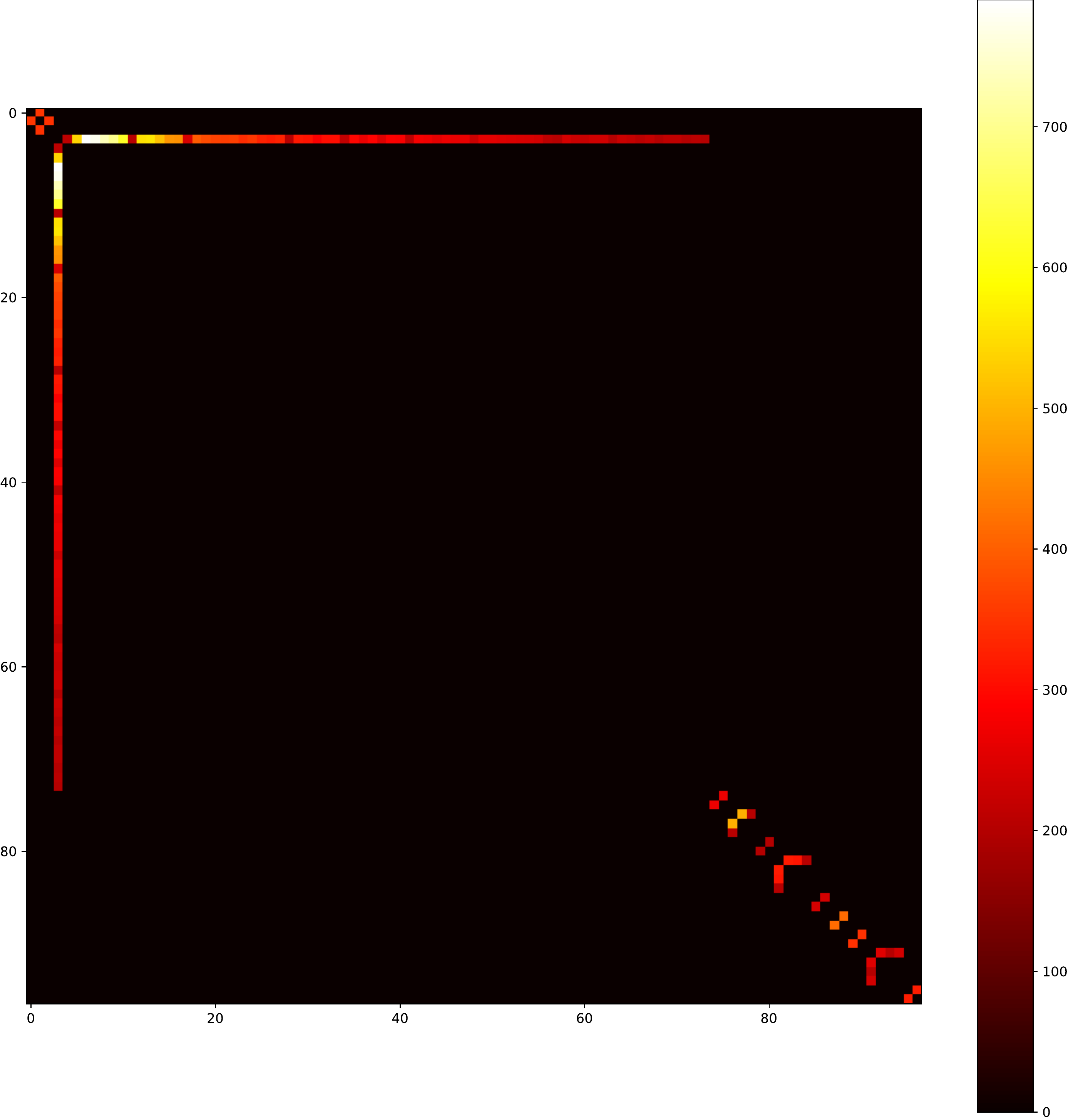}
\includegraphics [width=0.28\textwidth]{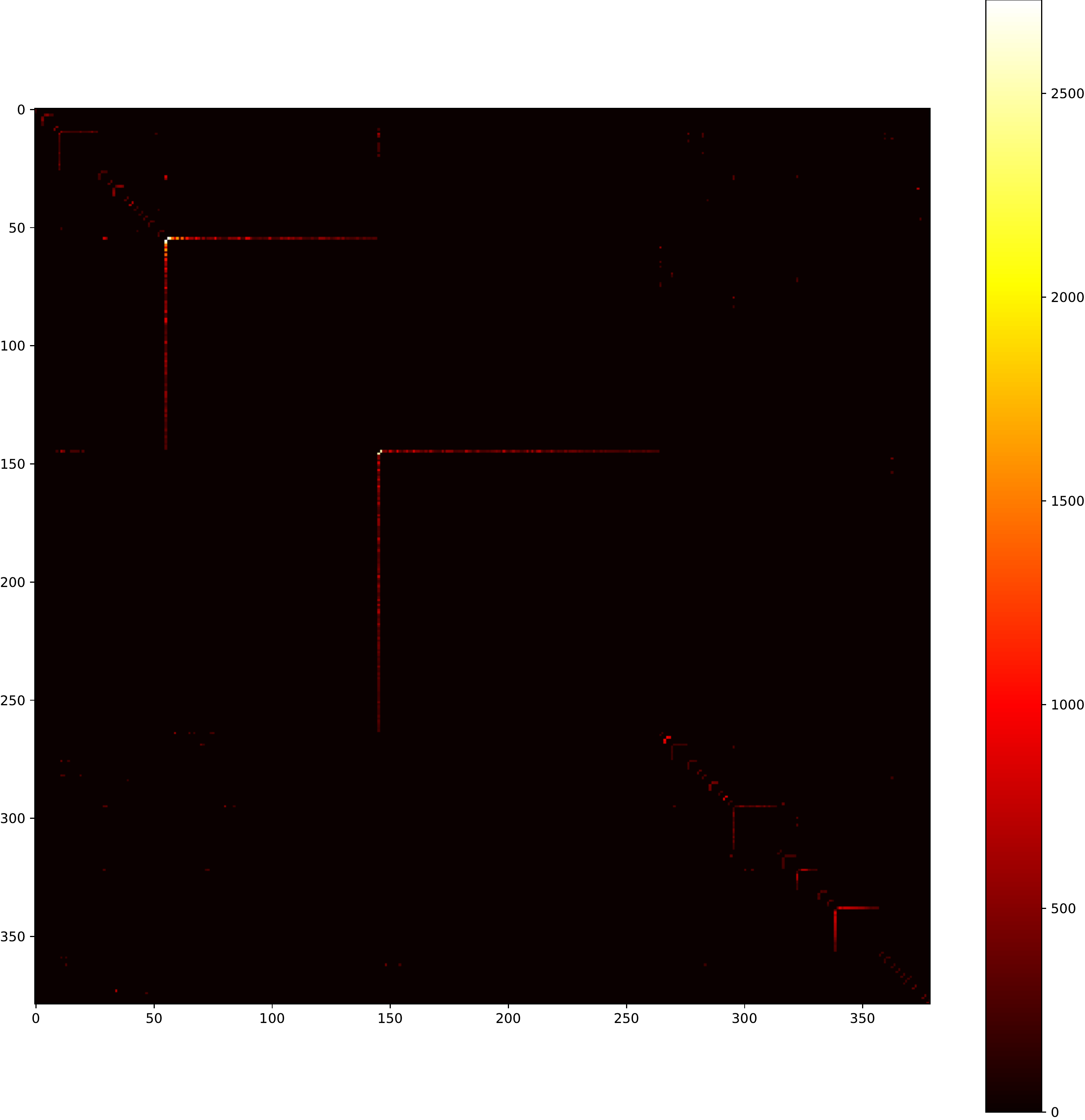}}
}
\caption{South American countries - Matrix Adjacency}
\label{fig:3_2}
\end{figure*}

The results for Argentina, Bolivia, Chile, Colombia, Ecuador, Paraguay, Peru, Uruguay and Venezuela, besides a color bar is showed to express the strength or weight of each edge, see Fig. \ref{fig:3_2}.

Analyzing the present results, it is possible identify network of users in each country. Besides, Bolivia is a country with less number of users therefore the filtered matrix is smaller than other countries. At the same time is possible to identify the line structures for Colombia, Ecuador, Peru, Paraguay, Uruguay and Venezuela. This lines can mean or represent a group of users with constant tagging between themselves. Venezuela calls our attention, considering the size of the filtered matrix and the extension of the lines, it is possible to identify two big groups of 100 and 150 users tagging themselves durin the period of study.

\subsection{Finding communities}

\begin{figure*}[hbpt]
\centerline{
\subfloat[Argentina, Bolivia, Chile]{
\includegraphics [width=0.3\textwidth]{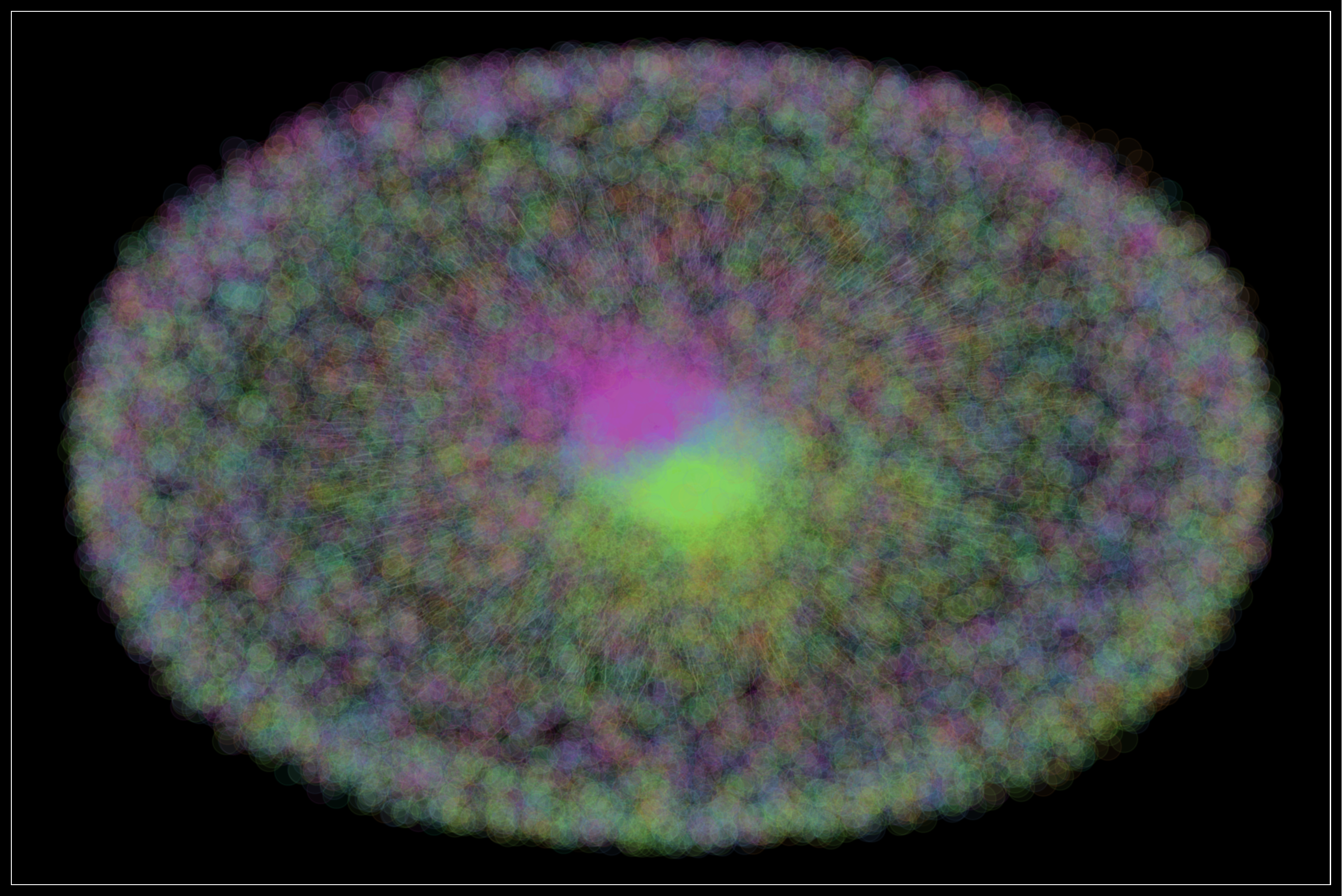}
\includegraphics [width=0.3\textwidth]{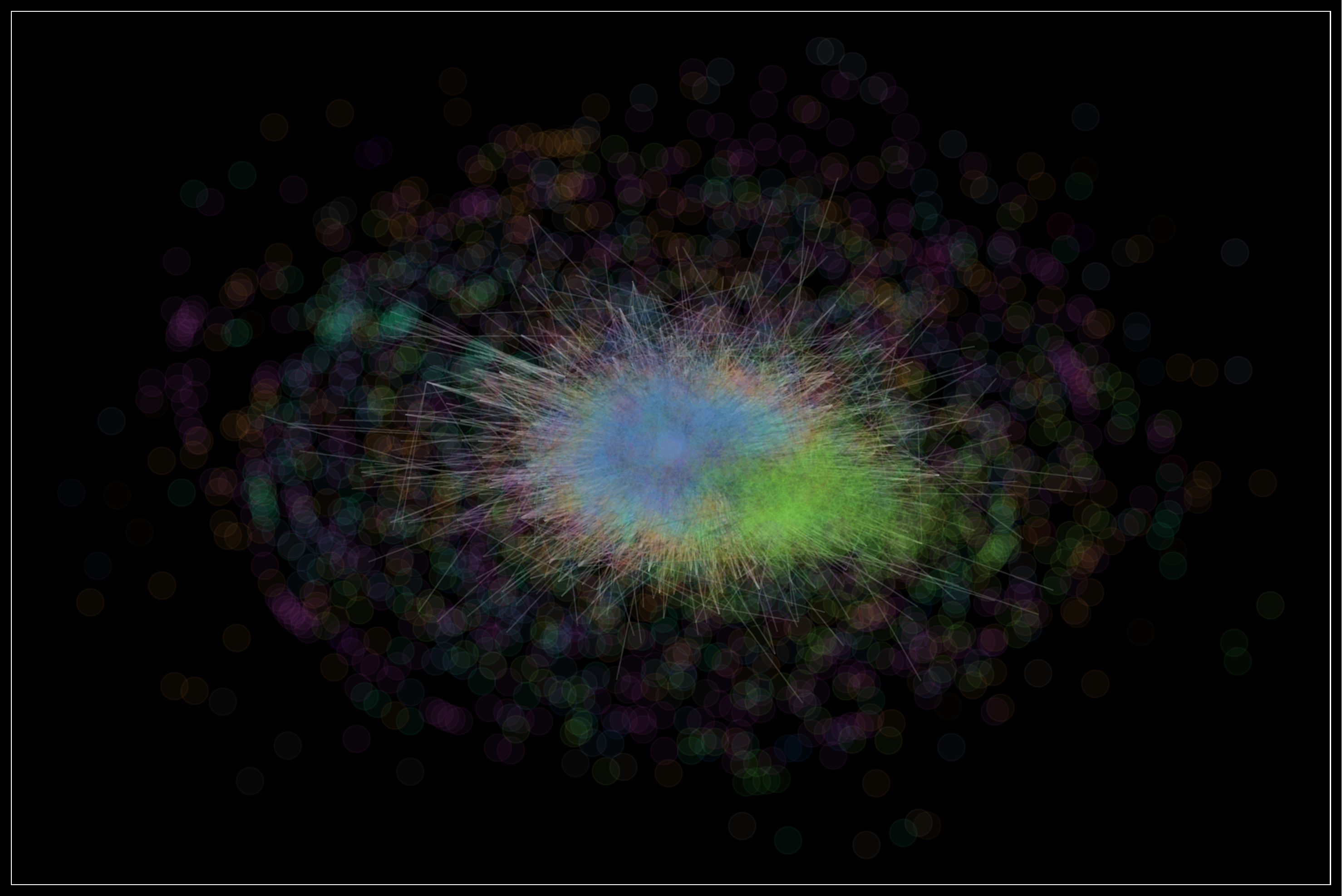}
\includegraphics [width=0.3\textwidth]{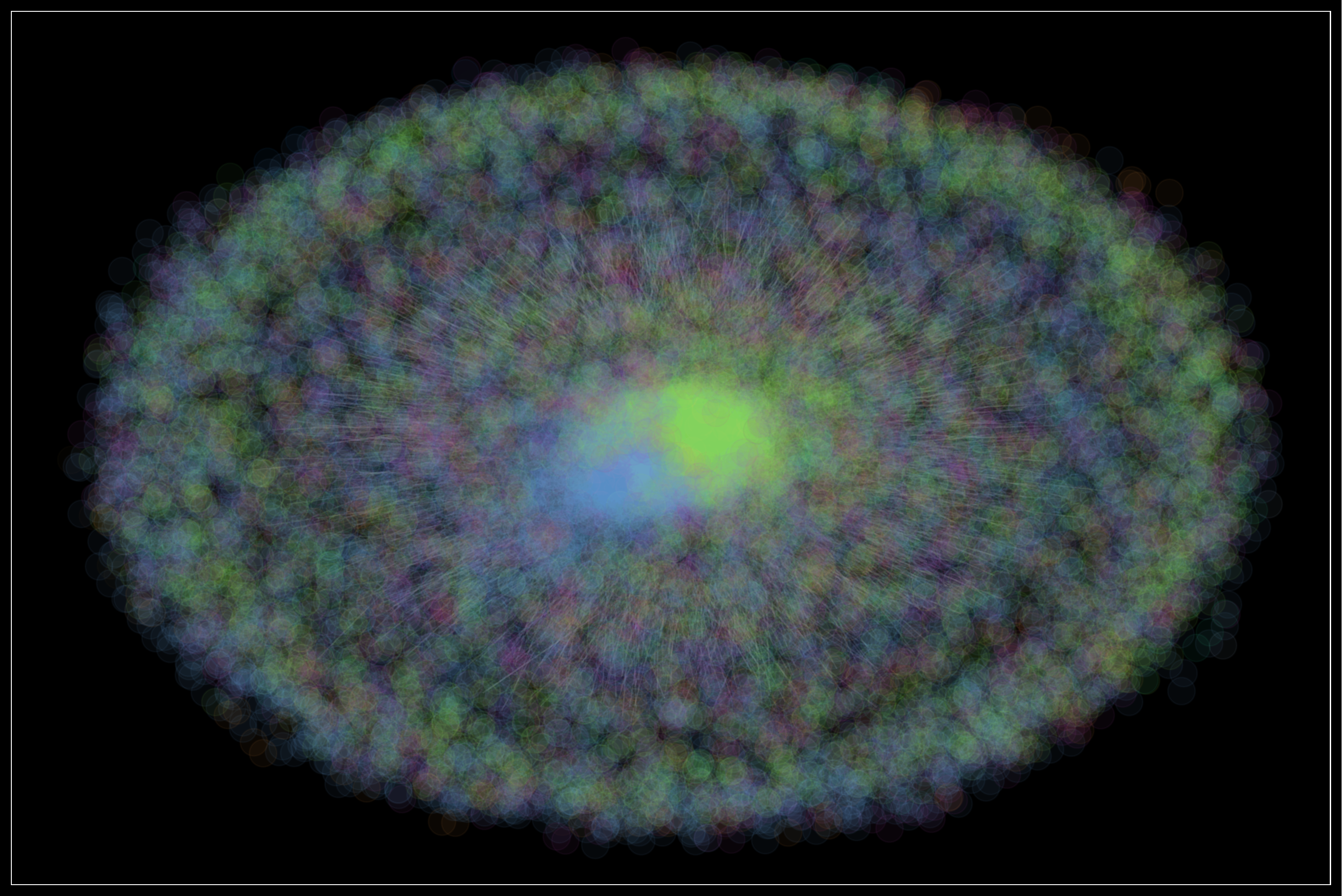}}
}
\centerline{
\subfloat[Colombia, Ecuador, Peru]{
\includegraphics [width=0.3\textwidth]{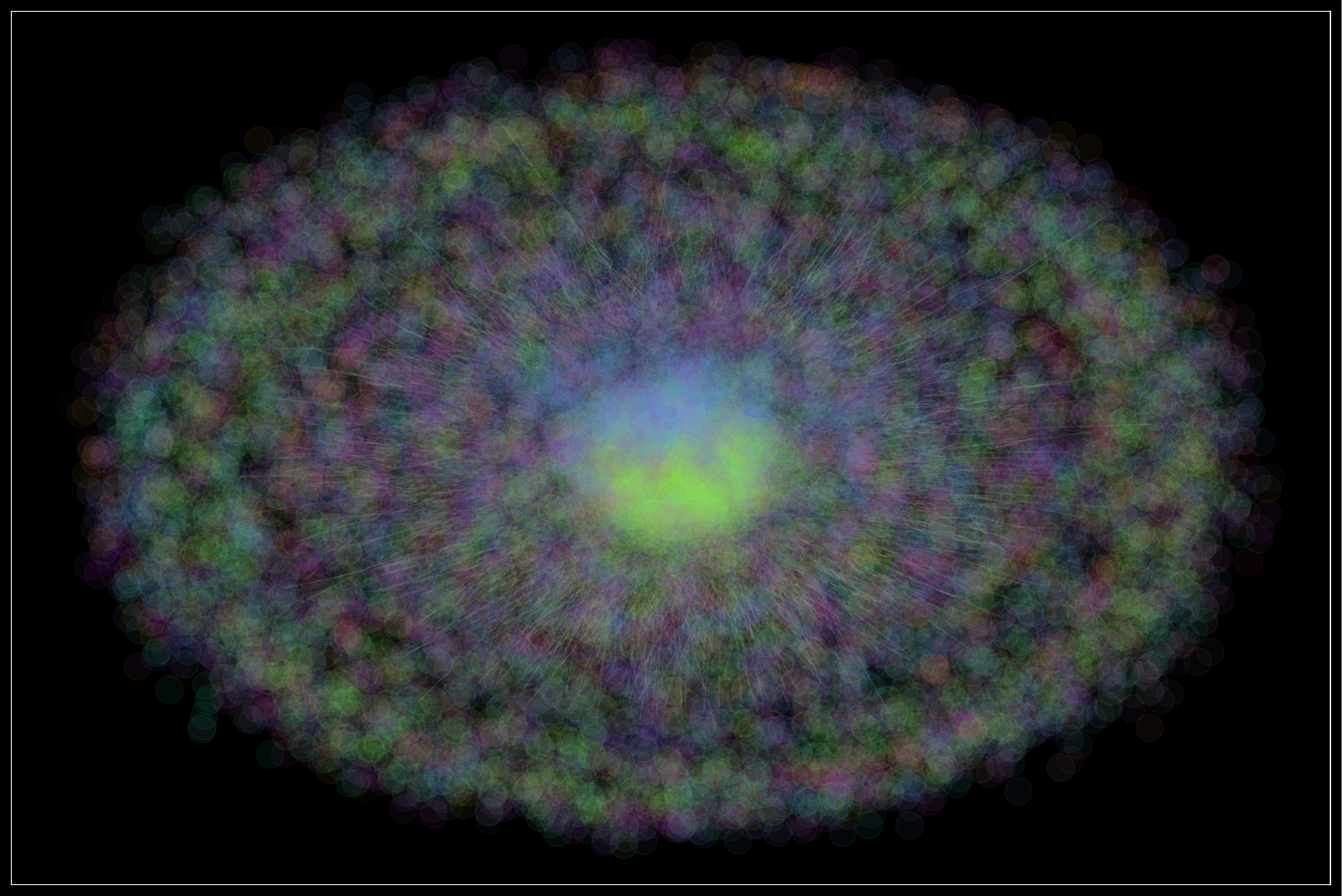}
\includegraphics [width=0.3\textwidth]{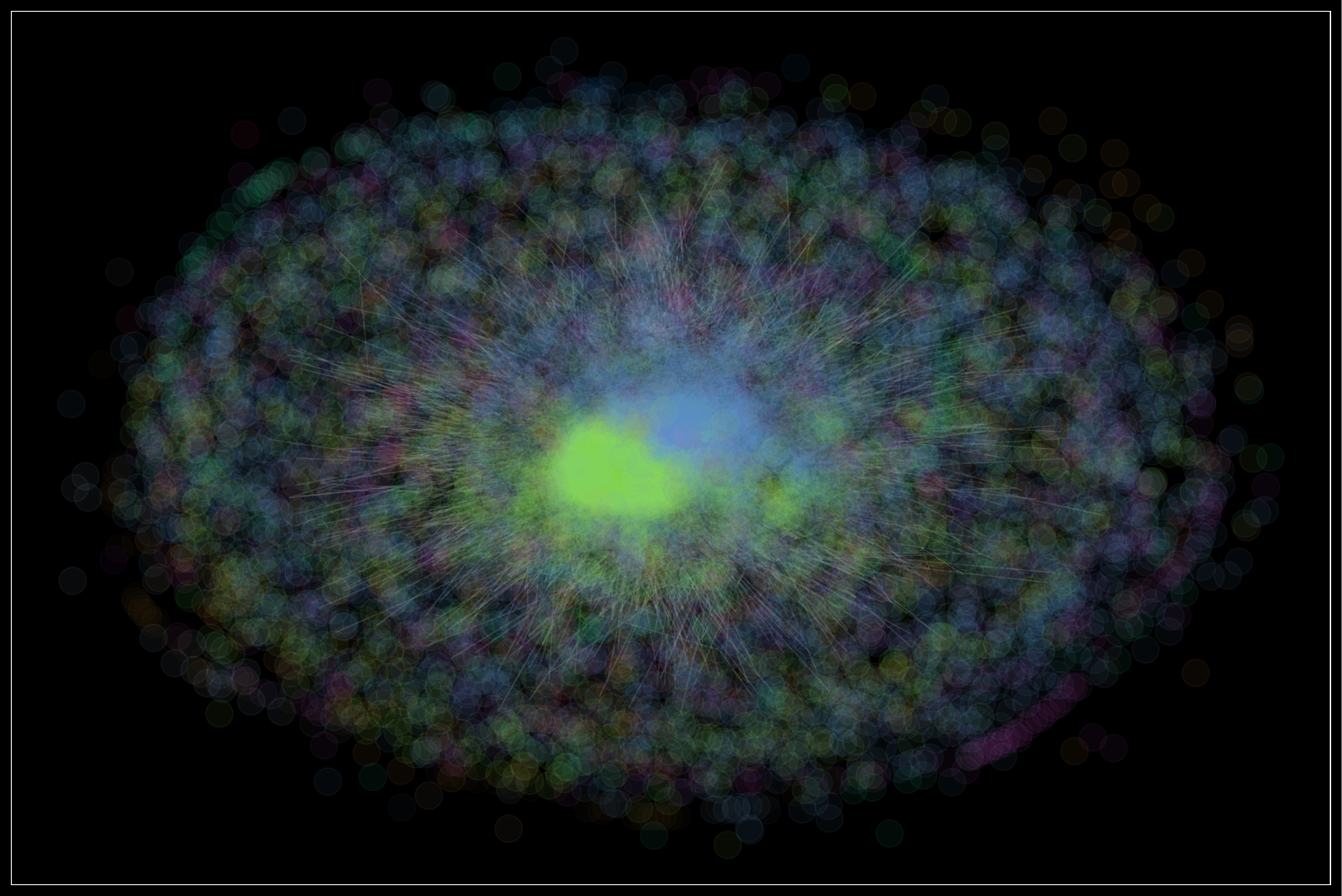}
\includegraphics [width=0.3\textwidth]{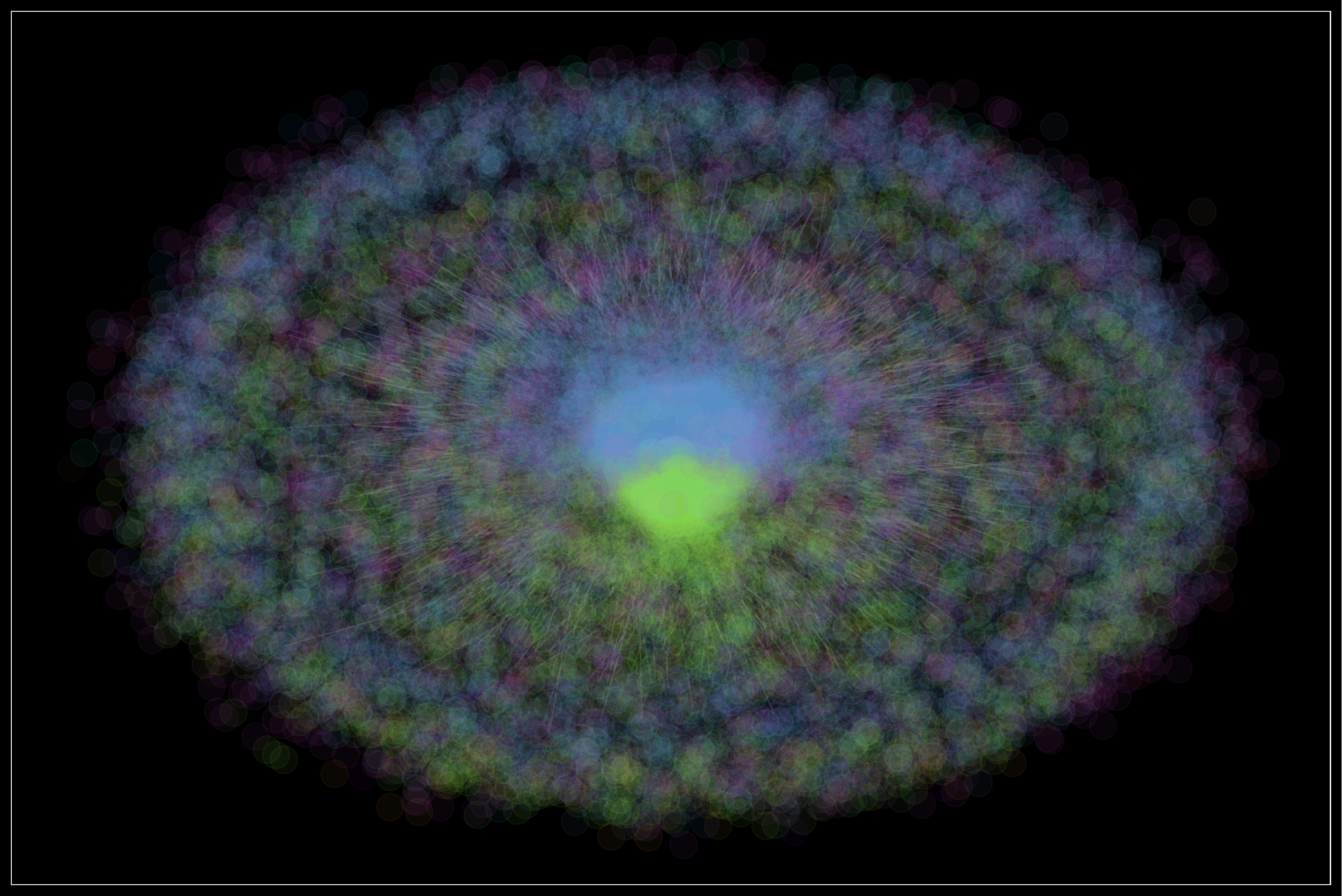}}
}
\centerline{
\subfloat[Paraguay, Uruguay, Venezuela]{
\includegraphics [width=0.3\textwidth]{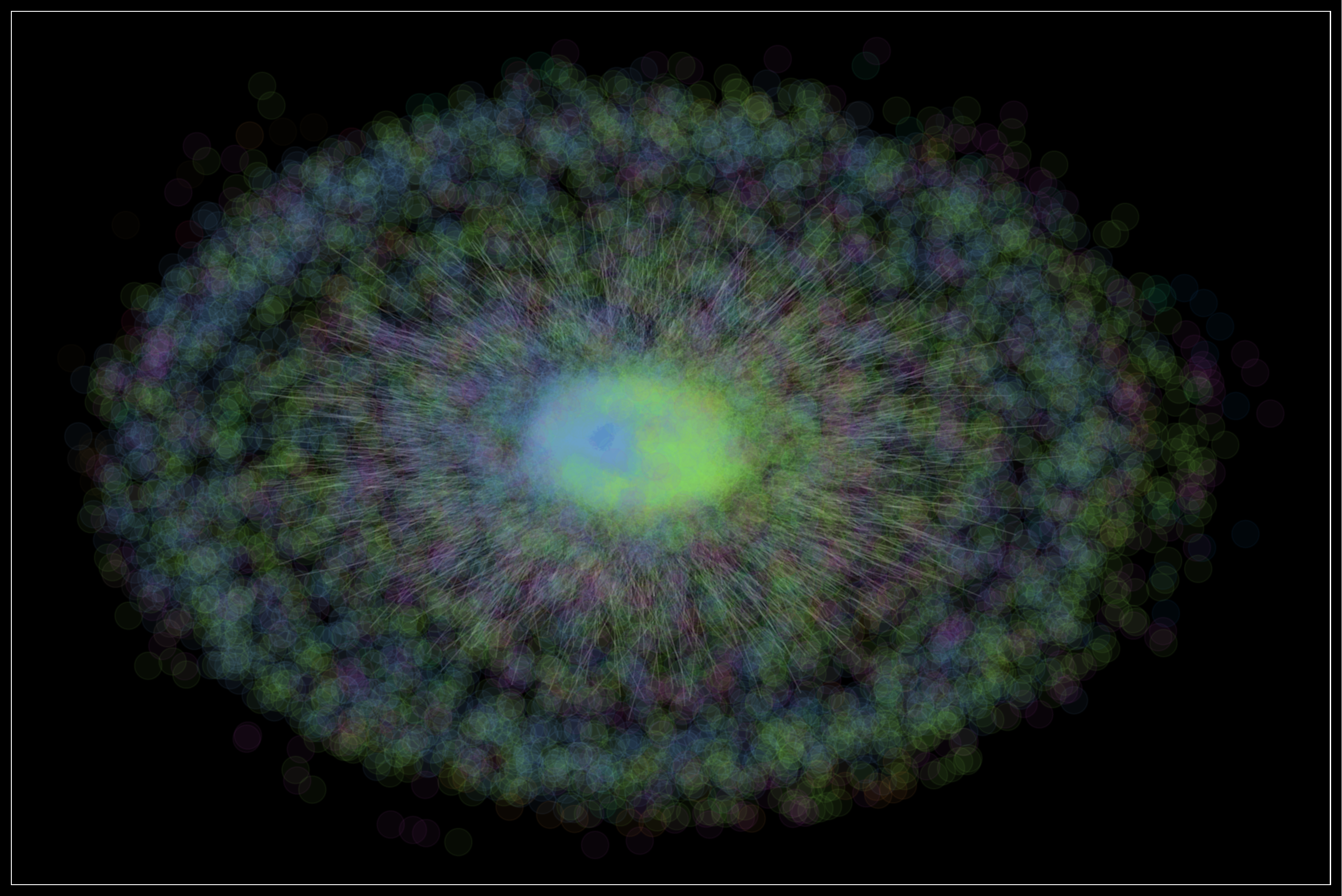}
\includegraphics [width=0.3\textwidth]{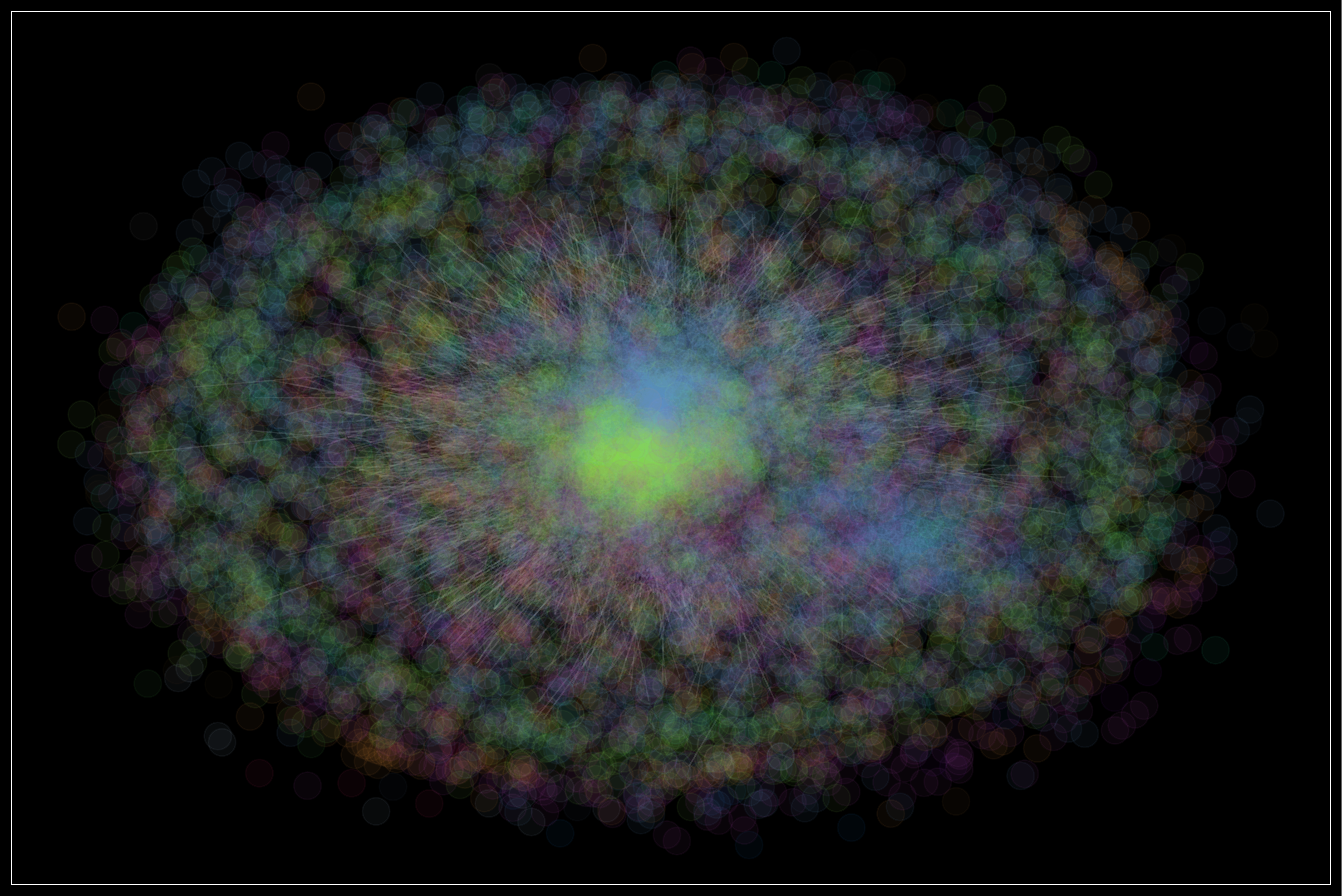}
\includegraphics [width=0.3\textwidth]{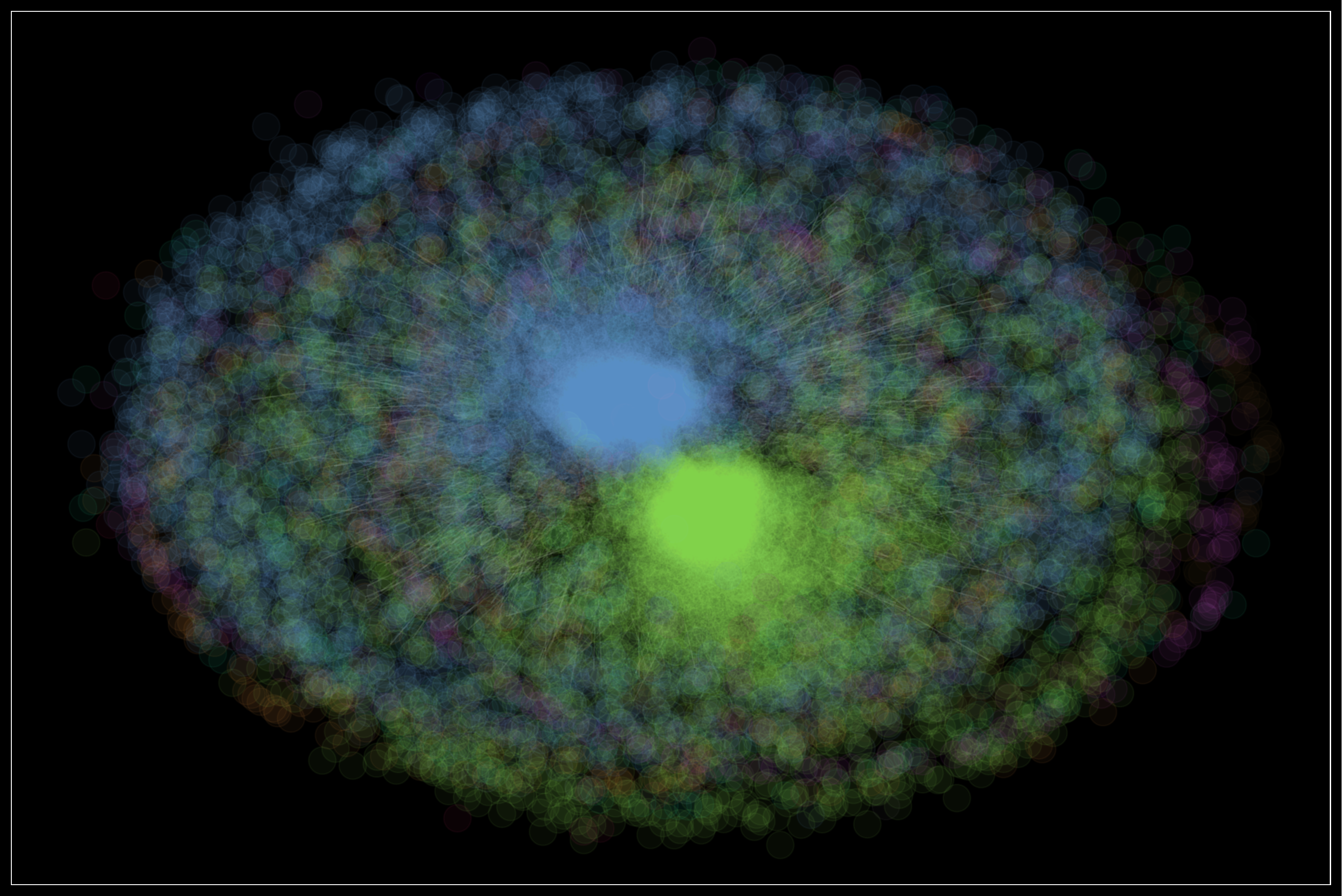}}
}

\caption{Communities}
\label{fig:3_3}
\end{figure*}

To conduct a deep analysis about the existence of users' groups an algorithm to find communities is performed over the complete graph. The results are presented in Table II and Figure \ref{fig:3_3}.

\begin{table}[]
\caption{Number of communities per country}
\centering{
\begin{tabular}{|c|c|}
\hline
          & \begin{tabular}[c]{@{}c@{}}Number of\\ communities\end{tabular} \\ \hline
Argentina & 34                                                              \\ \hline
Bolivia   & 39                                                              \\ \hline
Chile     & 19                                                              \\ \hline
Colombia  & 47                                                              \\ \hline
Ecuador   & 30                                                              \\ \hline
Peru      & 36                                                              \\ \hline
Paraguay  & 28                                                              \\ \hline
Uruguay   & 37                                                              \\ \hline
Venezuela & 26                                                              \\ \hline
\end{tabular}
}
\end{table}

A previous hypothesis about Venezuela can be confirmed considering the number of present communities in the interaction of users. And, the images representing the communities can express a big concentration of nodes in the center respectively showing the tagging of some specific users.

\section{Conclusion}
\label{sec:conclusion}

This preliminary study to analyze Twitter interaction of South American around covid-19 pandemic shows promissory results. First, a text mining approach is used to process text and find users. Second a proposal is performed in experiments sections showing the viability of creating Complex Network with the proposal. Finally, visualisation techniques are proposed to analyze the matrix adjacency of each country, a filtering process to select most representative behaviour and discovering of communities, Venezuela arises a concern about intentional group of users publishing content during this period of study. 

\section{Future Work}

Future work, involves to perform an analysis of every week with the aim of finding changes in the interaction, number of top users and describe all this behaviour using Complex Networks and features related to degree measurements of the nodes, edges.

\section*{Acknowledgments}
The author wants to mention Research4Tech, an Artificial Intelligence community of Latin American Researchers for promoting Science and collaboration in Latin American countries, his roles as integrator between Professional, Researchers, Technology communities is key to develop the Latin American region as a strong body.

\bibliographystyle{IEEEtran}
\bibliography{biblio.bib}

\end{document}